%% file: paper.tex
\pdfoutput=1
\documentclass[aps,twocolumn,showpacs,superscriptaddress,prb,10pt]{revtex4-1}
\usepackage{graphicx}
\usepackage{caption}
\usepackage{dcolumn}
\usepackage{bm}
\usepackage{color}
\usepackage{amsmath}
\usepackage{amsfonts}
\usepackage{subfig}
\usepackage[makeroom]{cancel}
\usepackage{natbib}

\newcommand{\COMMENTED}[1]{}

\begin{document}
\bibliographystyle{plainnat}
\author{Adam Chiciak}
\affiliation{Department of Physics, The College of William \& Mary, Williamsburg, Virginia 23187}

\author{Ettore Vitali}
\affiliation{Department of Physics, California State University, Fresno, Fresno, CA 93740}
\affiliation{Department of Physics, The College of William and Mary, Williamsburg, Virginia 23187}


\author{Shiwei Zhang}
\affiliation{Center for Computational Quantum Physics, Flatiron Institute, 162 5th Avenue, New York, New York 10010}
\affiliation{Department of Physics, The College of William and Mary, Williamsburg, Virginia 23187}






\title{Magnetic and charge orders in the ground state of the Emery model \\
 - accurate numerical results}

\begin{abstract}
We perform extensive auxiliary-field quantum Monte Carlo (AFQMC) calculations 
for the three-band Hubbard (Emery) model 
in order to study the ground-state properties of Copper-Oxygen planes in the cuprates. 
Employing cutting-edge AFQMC techniques with a self-consistent gauge constraint in auxiliary-field space to 
control the sign problem, we reach supercells containing 
$\sim500$ atoms to capture collective modes in the charge and spin orders and characterize the behavior in the thermodynamic limit.
The self-consistency scheme interfacing with generalized Hartree-Fock calculations allows high accuracy in AFQMC 
to resolve small energy scales, 
which is crucial for determining the complex candidate orders in such a system. 
We present detailed information 
on the charge order, spin order, momentum distribution, and localization properties as a 
function of charge-transfer energy for the the under-doped regime.
In contrast with the stripe and spiral orders under hole-doping, we find that the corresponding $1/8$ electron-doped system
exhibits purely antiferromagnetic order in the three-band model, consistent with the asymmetry between electron- and hole-doping
in the phase diagram of cuprates.

\end{abstract}

\maketitle

\section{Introduction}
\label{sec:intro}


Significant progress has been made in the study of a variety of strongly correlated electron
systems\cite{Tranduada_AIP,Tranquada_stripes,Fujita_stripes,Chang_nature,Fischer_nematic,Poilblanc_HF,Zaanen_Gunnarsson,Millis_stripes,PhysRevB.92.205112,White_Scalapino_stripes,Sarker_spiral,Thomson_Sachdev,White_checkerboard,RevModPhys.66.763}. 
However, despite more than thirty years of theoretical and experimental studies,
major questions remain in understanding high--temperature superconductivity. 
Recent advances in computing technology and computational methods are providing new
opportunities to address important questions with more powerful and more systematic
computational studies.  

It is widely believed that the superconducting order of the cuprates arises from a physical 
mechanism in the quasi--two--dimensional planes hosting the copper and oxygen atoms\cite{Emery}. 
Other layers of the material play the role of charge reservoirs, which can be used to dope
the copper--oxide planes by adding or removing electrons (holes).
The experimental evidence 
indicates that, when no doping is present, in the
parent compounds, the stable phase is an insulating antiferromagnet\cite{Armitage_Mott}.
With doping, this order rapidly disappears, 
giving rise to a rich, complicated phase diagram
with respect to doping and temperature in which different spin and charge orders appear to coexist,
either cooperating or competing\cite{RevModPhys.87.457, PhysRevLett.94.156404}. 
From the theoretical point of view, the complex electronic structure of these materials makes
a fully {\it{ab initio}} many-body computation 
a formidable task, in particular since the characterization
of the phases requires a detailed study of the bulk limit.
Because of this, 
a main focal point of the theoretical research is 
creating minimal models to study the order in the copper--oxide plane.

Most of the effort to model this problem has focused on the Hubbard Hamiltonian\cite{Anderson} (and the related $t$-$J$ model).
This model relies on the Zhang--Rice singlet notion\cite{PhysRevB.37.3759} that allows 
a reduction of the degrees of freedom by 
treating the oxygen sites implicitly in the mathematical description.
A variety of 
accurate numerical 
results have been obtained for the one-band Hubbard model,
which  for example
indicate the existence of 
stripe and spin-density wave orders in the ground state \cite{Mingpu-sc-PRB,Hubbard-benchmark,science-17},
compatible with those observed in experiments on the real materials, although quantitative agreement is not always achieved.
Perhaps more importantly, results from the one-band model show
the close and delicate competition between 
different orders consistent with experimental observations.
Indications are, however, that the pure Hubbard model (no hopping beyond nearest neighbor) does not appear to display a
superconducting ground state at intermediate coupling and reasonable doping 
 \cite{Mingpu-pure-Hubbard-SC}.  
This gives more impetus to look more closely beyond the simplest models.

Recent X-ray scattering experiments and nuclear magnetic resonance experiments 
indicate that the oxygen $p$-bands are 
involved in spin and charge
density wave states \cite{Comin_xray,Jurkutat_NMR,Rybicki2016,Achkar_Xray,Haase_NMR}. 
This suggests one direction to improve the model may be to include
the oxygen $p$-bands explicitly as non-trivial hole carriers. With recent advances in computational 
methodologies, 
several sophisticated many-body approaches can now go well beyond the minimal 
Hubbard model to study the 
more realistic 
three-band Hubbard model, or Emery model\cite{Emery}.
This model explicitly includes copper $3d_{x^2-y^2}$
and the oxygen $2p_x$ and $2p_y$ orbitals. The model Hamiltonian contains several parameters,
including the charge transfer energy, hopping amplitudes and on-site repulsion 
energies for the different bands. 

In principle  the Hamiltonian parameters can be computed from approximate {\it{ab initio}} approaches. However their
actual determination is subtle. 
In particular the value of the charge transfer energy, 
which carries the physical meaning of the energy required to move a hole from a copper $d_{x^2-y^2}$
orbital to a oxygen $p$ orbital, 
$\Delta$, 
can be affected by double--counting issues\cite{Wang_double} in the computation.
In addition, the value of the charge transfer energy varies across the different families
of the cuprates, and it controls the average electron occupation around copper and oxygen
atoms. 
There are indications that the charge--transfer energy and, in effect, the average copper occupation are anti--correlated with 
the critical temperature\cite{RUAN20161826,Jurkutat_NMR,Rybicki2016,0295-5075-100-3-37001}, 
which makes $\Delta$ a crucial parameter in the Emery model. 
From a recent auxiliary-field quantum Monte Carlo study of the model at half-filling 
\cite{Ettore_three-band}, 
we have seen that the properties of the ground state of the model vary 
fundamentally with $\Delta$, 
showing a quantum phase transition transition from an insulating antiferromagnet to a non--magnetic metal.

Away from half-filling, there have been many computational studies addressing the behavior of the Emery model
using different methodologies, including 
exact diagonalization of small clusters\cite{Dobry_ED}, 
random phase approximation \cite{Atkinson_RPA}, 
Density Matrix Renormalization Group  \cite{PhysRevB.92.205112}, 
quantum Monte Carlo\cite{Dopf_QMC, Scalettar_QMC,Yanagisawa_3bandQMC,CPMC_3band}, 
and embedding methodologies (including dynamical mean field theory and cluster extensions, and density matrix embedding theory)
\cite{Arrigoni_cluster, PhysRevB.78.035132,0295-5075-100-3-37001,Garnet-group-3-band}.
%
These calculations have revealed a great deal about the properties of the model. Many similarities are seen between this model and 
the one-band Hubbard model, including the presence of strong magnetic correlations 
away from half-filling and their delicate balance or competition with 
superconductivity. 
Even so, the numerical evidence has been inconclusive on several key issues, including the nature of the true ground state,
because of 
computational limitations 
including the accuracy of the many-body methods. 
This is not surprising, given that  even in the one-band model some of these issues are 
only now being resolved using combinations of the latest advances in computational methodologies.

One of the challenges in characterizing the ground-state magnetic and charge order is the difficulty that most numerical methods face in 
reaching both the thermodynamic and zero-temperature limits. 
In a recent study, Huang et al \cite{Huang-fluctuating-stripes}  
found the presence of fluctuating stripes in  the model at high temperature.
We have recently carried out a Generalized Hartree-Fock study \cite{Chiciak_GHF} of the magnetic and charge orders 
in this model, focusing on the ground-state phase diagram and its dependence on $\Delta$. The results indicate 
the existence of long-wavelength collective modes as was seen in the one-band Hubbard 
model \cite{science-17,chia-chen-prl-10,PhysRevB.95.125125-Hub-DMRG-stripe}. 
In addition to stripes, they also suggested possible additional orders as $\Delta$ is varied, such as spirals and magnetic domains. 
Such states are extremely challenging to detect and establish, because of the requirements on both accuracy and robustness of the underlying 
many-body method and the capability to reach large system sizes to approach the thermodynamic limit.


Motivated by these considerations and to understand how similar or different this model is from the  one-band Hubbard model,
we investigate 
the three-band Hubbard model 
at zero-temperature,
using 
state-of-the-art auxiliary-field quantum Monte Carlo (AFQMC) algorithm. 
We focus on the nature of the spin and charge orders, and seek to quantify the competition between stripes,  spin spirals, and other nematic orders
in the thermodynamic limit.
We provide accurate, detailed numerical results on the ground state in the underdoped regime. At larger $\Delta$ roughly consistent with 
the Y-based family, we find stripe order at $1/8$-hope doping. At smaller $\Delta$ where the $d$- and $p$-orbital occupancies are more in line with 
 Hg-, Bi-, and Tl-based cuprate families, we find close competition between stripe and spiral states. 
Although most of our results are for the underdoped regime, we find that the $1/8$ electron-doped case larger $\Delta$  shows a tendency for 
phase-separated antiferromagnetic (AFM) order, in contrast with the hole-doped case.
Our method employs a self-consistent constraint \cite{Mingpu-sc-PRB} on paths sampled in auxiliary-field space, which has been referred to as the constrained path (CP) 
approximation
when applied to many-body model systems  \cite{CP-review-lecture-notes-2019}. The CP approach relies on a trial wave function or density matrix 
for a sign or gauge condition 
 on 
the sign or gauge of the Slater determinants sampled in the AFQMC, thereby controlling the sign problem.
This approach has consistently demonstrated 
a high level of accuracy and allowed robust predictions in the one-band Hubbard model \cite{Hubbard-benchmark,science-17,chia-chen-prl-10}.

The rest of the paper is organized as follows. In Sec.~\ref{sec:model}, 
we introduce the three-band Hubbard model. 
In Sec~\ref{sec:method}, we briefly describe the CP AFQMC method as
well as the self--consistent scheme used.
In Sec.~\ref{sec:results}, we present our findings for the 
(\ref{ssec:spin-charge}) spin and charge order,
(\ref{ssec:momentum}) momentum distributions,
(\ref{ssec:exotic}) localization properties. 
and (\ref{ssec:asymmetry}) the hole-electron asymmetry.
We further discuss results and make conclusions
in Sec.~\ref{sec:conclusion}.

\section{Model}
\label{sec:model}

\begin{figure}[ptb]
\includegraphics[width=0.6\columnwidth, angle=0]{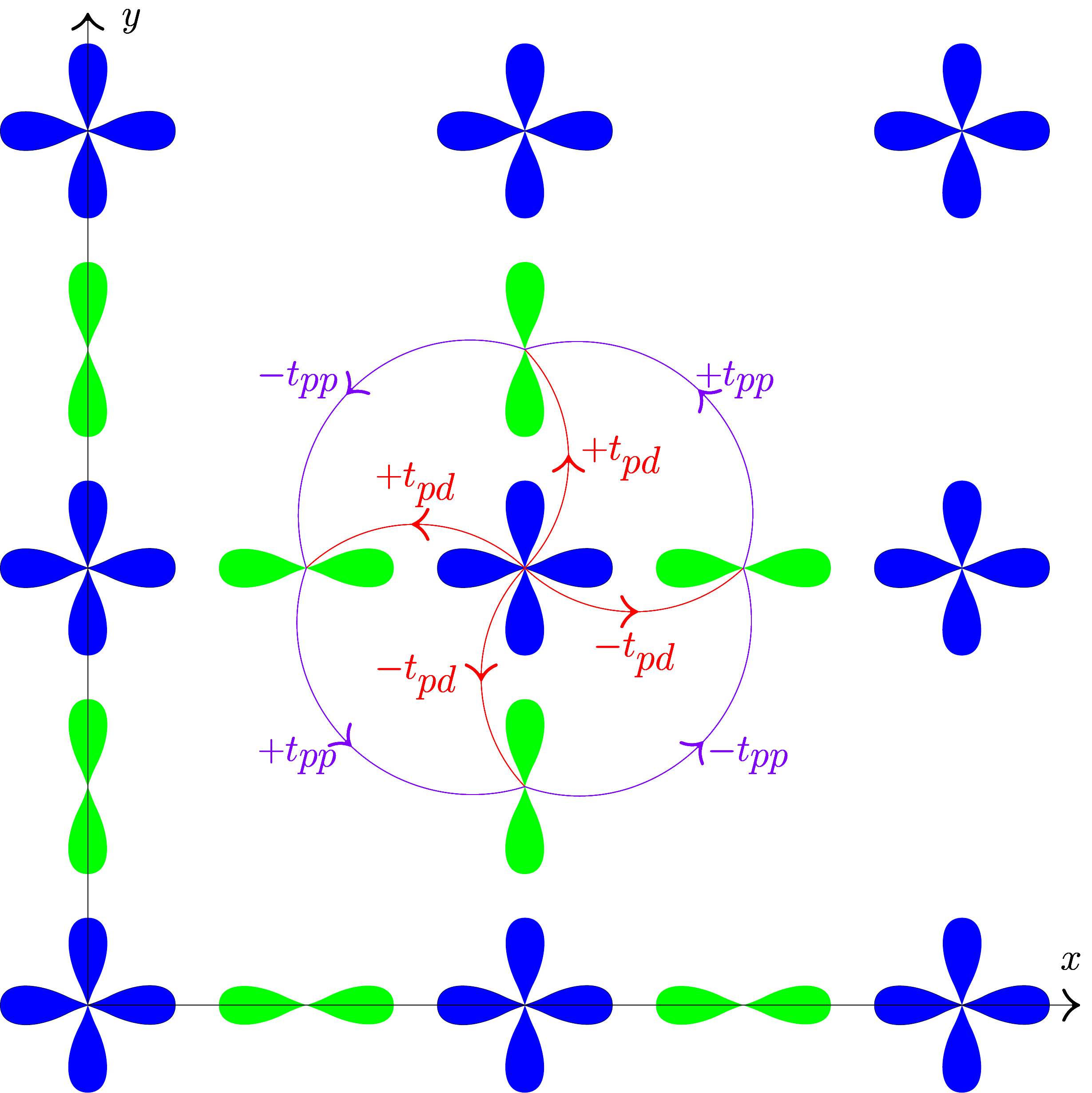}
\caption{
 (Color online) Schematic view of the CuO$_2$ planes in cuprates 
 and illustration of the 3-band model.
 Cu $3d_{x^2 - y^2}$ orbitals are represented in blue,
 and O $2p_x$ and $2p_y$ orbitals in green.
 We use the reference frame defined by the two axes in the figure.
 The curve connectors represent the hopping, and the labels
 define the sign rule.}
\label{fig:model}
\end{figure}

The Emery model, also called three-band Hubbard model, includes the Cu $3d_{x^2 - y^2}$ orbital and 
the O $2p_x$ and $2p_y$  orbitals explicitly in the description of the copper-oxide planes in the cuprates.  
In Fig.~\ref{fig:model}, 
a schematic representation of one CuO$_2$ plane is shown to help visualize the model.
We will consider simulation supercells made of $M=L_x \times L_y$ unit cells of CuO$_2$, with
a given number of particles (or more precisely of holes), $N$ , which then defines the density or doping.
The Hamiltonian is 
\begin{equation}
\label{3bands:ham}
\begin{split}
& \hat{H} = \varepsilon_d \sum_{i, \sigma} \hat{d}^{\dagger}_{i,\sigma} \hat{d}^{}_{i,\sigma} 
+ \varepsilon_p \sum_{j, \sigma} \hat{p}^{\dagger}_{j,\sigma} \hat{p}^{}_{j,\sigma} + \\
& \sum_{<i,j>, \sigma} t_{pd}^{ij} \left(\hat{d}^{\dagger}_{i,\sigma} \hat{p}^{}_{j,\sigma} + h.c \right)
+  \sum_{<j,k>, \sigma} t_{pp}^{jk} \left(\hat{p}^{\dagger}_{j,\sigma} \hat{p}^{}_{k,\sigma} + h.c \right)
\\
& + U_d  \sum_{i} \hat{d}^{\dagger}_{i,\uparrow} \hat{d}^{}_{i,\uparrow} \hat{d}^{\dagger}_{i,\downarrow} \hat{d}^{}_{i,\downarrow} + U_p \sum_{j} \hat{p}^{\dagger}_{j,\uparrow} \hat{p}^{}_{j,\uparrow} \hat{p}^{\dagger}_{j,\downarrow} \hat{p}^{}_{j,\downarrow}\,.
\end{split}
\end{equation}
In Eq.~\eqref{3bands:ham}, $i$ runs over the sites $\vec{r}=(x,y)$ of a square lattice
$\mathbb{Z}^2$ defined by the positions of the Cu atoms, $\vec{r}_{\rm Cu}$.
The labels $j$
and $k$ run over the positions of the O atoms, shifted with respect
to the Cu sites, $\vec{r}_{{\rm O}_x} = \vec{r}_{\rm Cu} + 0.5 \, \hat{x}$ for the $2p_x$ orbitals, and
$\vec{r}_{{\rm O}_y} = \vec{r}_{\rm Cu} + 0.5 \, \hat{y}$ for the $2p_y$ orbitals. 
The model is formulated in terms of holes rather than electrons: for example, the operator $\hat{d}^{\dagger}_{i,\sigma}$
creates a hole on the $3d_{x^2 - y^2}$ orbital at site $i$ with spin $\sigma = \uparrow,\downarrow$.
The first two terms in the  
Hamiltonian contain the orbital energies, which define the charge-transfer
energy parameter $\Delta \equiv \varepsilon_p - \varepsilon_d$, which can be thought of as the energy needed for a hole
to move from a Cu $3d_{x^2 - y^2}$ orbital to an O $p$ orbital.
The next two terms  
describe hopping between orbitals; the hopping amplitudes $t_{pd}^{ij}$ and $t_{pp}^{jk}$ are expressed in terms
of two parameters,  
$t_{pd}$ and $t_{pp}$,
and the dependence on the sites is simply a sign factor,  
as depicted in  Fig.~\ref{fig:model}.
Finally, the last two terms 
represent the on-site repulsion energies, or
double-occupancy penalties, similar to those in the one--band Hubbard model. We neglect Coulomb interactions beyond the on-site terms.

We study the properties of the model as a function
of the charge transfer energy $\Delta$.
Our starting point is an {\it{ab intio}} set \cite{Wagner_abinitio} of parameters obtained for La$_2$CuO$_4$, the parent compound of
the lanthanum based family of cuprates.
The parameter values are listed in Table~\ref{tab:param_table}.
This set corresponds to a charge transfer energy $\Delta = 4.4$ eV.
To correct for possible double counting issues \cite{Wang_double} would imply 
a considerable reduction of this value
to $\Delta \sim 1.5$ eV, which as pointed out above, can greatly change the physics of a system.

\begin{table}
\begin{center}
\caption{Parameter values adopted in the present study. 
The parameters are obtained from La$_2$CuO$_4$  \cite{Wagner_abinitio}. 
We study the value of 
$\Delta = \varepsilon_p - \varepsilon_d$ at $4.4$ and $2.5$.
}
  \begin{tabular}{ l c  c  c  c c c}
    \hline\hline
      Parameter & $U_d$ & $U_p$ & $\varepsilon_d$ & $\varepsilon_p$ & $t_{pd}$ & $t_{pp}$ \\
       \hline
	Value (eV) & 8.4 & 2.0 & -8.0 & -3.6 & 1.2 & 0.7\\
	\hline\hline
  \end{tabular}
\label{tab:param_table}
\end{center}
\end{table}

Most of our  calculations are performed at hole-doping, $h = 1/8$. The hopping
and on-site interaction parameters are kept at the values 
given in Table~\ref{tab:param_table},
and the charge--transfer energy, $\Delta$, is varied. Building on our half-filling study \cite{Ettore_three-band}, we focus on
two particular values, $\Delta = 4.4$ and $2.5$, which are representative of the insulating
and conducting states at half--filling, respectively. 

\section{Methods}
\label{sec:method}

To compute the ground state properties of the model in Eq.~\eqref{3bands:ham}
for a given system, i.e., a given set of parameters $(\varepsilon_d,\varepsilon_p,\{t^{ij}_{\alpha\beta}\},U_d,U_p )$ and supercell, we use 
the Constrained Path Auxiliary Field Quantum Monte Carlo (CP-AFQMC) method \cite{CPMC-PRB97,CP-review-lecture-notes-2019}.
In addition to tests in lattice models  \cite{Hubbard-benchmark}, this method has been shown in 
 a variety of other correlated systems to be
among the most accurate, low-polynomial scaling many-body methods 
 \cite{Motta-H-benchmark,Williams-etal-PRX-TMObenchmark}.

In order to sample the ground state $|\Psi_0\rangle$ of the Hamiltonian in Eq.~\eqref{3bands:ham}  
for a given supercell,
the technique relies on the imaginary-time evolution of an approximate initial wave function, say $|\psi\rangle$:
\begin{equation}
\label{time_projection}
|\Psi_0\rangle \propto \lim_{\beta\to +\infty} \exp(-\beta(\hat{H} - E_0))|\psi\rangle
\end{equation}
where  $E_0$ is the ground state energy which is estimated adaptively in the process.
The projection formula in Eq.~\eqref{time_projection} is valid for any 
$\langle \psi \, | \,\Psi_0\rangle \neq 0 $.
In the CP-AFQMC  algorithm, the imaginary-time evolution 
is mapped on to open-ended branching random walks in the manifold of Slater determinants, known as
the ``walkers.'' The sign problem is controlled 
through the introduction of a trial wave function, 
$|\psi_T\rangle$, which guides the random walks and imposes a sign constraint 
by eliminating random walk paths when the 
overlap of a walker with $|\psi_T\rangle$  first turns negative. 
(A gauge constraint is applied on the overall phase of the Slater determinant in the case of walkers
described by Slater determinants with complex orbitals \cite{Phaseless-AFQMC}.)

In this study, we are concerned with the cooperating or competing magnetic and charge orders that may arise
in the three--band model as a function of the charge transfer energy. 
We define the spin 
on the Cu sites for the $d$-bands as
\begin{equation}
\hat{\mathbf S}({\mathbf r}) = \frac{1}{2} \sum_{\sigma,\sigma^{\prime}}
{\bm \sigma}_{\sigma,\sigma^{\prime}} \, \hat{d}^{\dagger}_{i,\sigma}
\hat{d}_{i,\sigma^{\prime}}\,,
\label{eq:def-S-Cu}
\end{equation}
where ${\bm \sigma}_{\sigma,\sigma^{\prime}}$ denotes the elements of the Pauli spin matrices. 
As in Eq.~\eqref{3bands:ham}, the label $i$ has a one-to-one correspondence with the 
position ${\mathbf r}=(x,y)$.
The spins on the O $p$-bands can be similarly written down, but they turn out to be negligible as we discuss below.
The charge densities are defined as 
\begin{equation}
{\hat{n}_\alpha}({\mathbf r}) = \sum_{\sigma} \, \hat{\alpha}^{\dagger}_{i,\sigma}
\hat{\alpha}_{i,\sigma}\,,
\label{eq:def-n}
\end{equation}
where $\alpha$ is either $d$ or $p_x$ or $p_y$, and 
the operator 
$\hat{\alpha}^{\dagger}_{i,\sigma}$ is the corresponding creation operator for
 a hole of spin $\sigma$ in the unit cell $i$.

In order to optimize the numerical 
detection of complex spin and charge orders, we
explicitly break translational and $SU(2)$ symmetry through the application of a
weak pinning field coupled to the local spin density on one side of the system:
\begin{equation}
\label{eq:pinning}
\hat{V}_{\rm ext}= \sum_{{\mathbf r}=(x,y)} \delta_{y,0} (-1)^{x+y}
 \, {\bf{h}}_{\rm pinn} \cdot \hat{\mathbf S}({\mathbf r}) 
\end{equation} 
where ${\bf{h}}_{\rm pinn} = (h_x,h_y,h_z)$ can be tuned to obtain the desired external field.
Throughout this paper, we choose the pinning field to be coupled to the in-plane $x$-component of the spin density, unless stated otherwise. 
This field induces a local AFM 
order on the $d_{x^2 - y^2}$ orbitals 
on one side of the system. The presence of the long-range order is determined by measuring the behaviors of the spin and charge density, 
after extrapolation of the results to the thermodynamic limit and to the $h_{\rm pinn} \to 0$ limit.  
The symmetry-breaking pinning field allows us to measure densities as opposed to correlation functions 
which would be needed in fully periodic calculations. 
This dramatically improves our resolution, since 
 at large distance (from the location of the pinning field)
  the order being numerically measured becomes 
$\mathcal{O}(S)$ rather than $\mathcal{O}(S^2)$, where $S$ is the ``order parameter'', for example, the magnitude of the spin. 

\subsection{Self-Consistent Constraint}
\label{ssec:self-consistent}     

\begin{figure}[ptb]
\includegraphics[width=\columnwidth, angle=0]{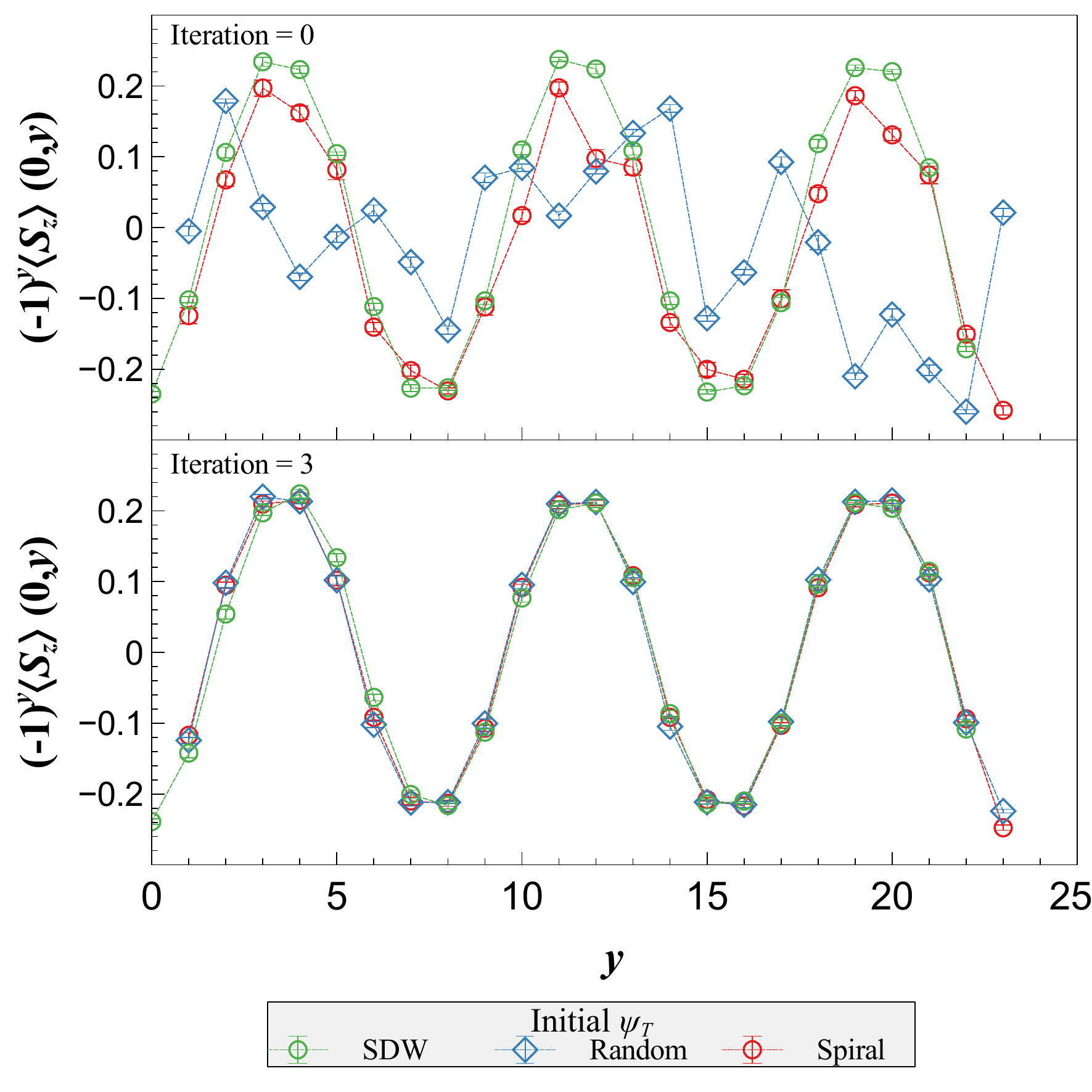}
\caption{
  (Color online) Convergence of the self-consistent constraint in AFQMC. 
  The $S_z$-component of the staggered spin vector along
  the line cut at $x = 0$, for three separate self-consistent calculations starting from varied
  initial states (SDW, Random, and Spiral).
The system is $4 \times 24$,
  at $\Delta = 4.4$, with PBC in the $x$-direction, an open boundary condition in the $y$-direction, and a pinning field in the 
  $z$-direction applied at $y=0$.
  The top panel is the spin order computed by AFQMC at the end of the $0^{th}$ iteration,
  i.e., using the initial state as trial wave function.
  The bottom panel shows that the order is converged by the $3^{rd}$ iteration. The final order is a linear SDW with
  the majority of the spin vector in the $S_z$-direction.
  }
\label{fig:convergence}
\end{figure}

The CP constraint  is an approximation which results in a systematic bias. The magnitude of the bias has been shown to be usually very small, even
with simple mean-field $|\psi_T\rangle$.
(For example, in the one-band Hubbard model with $U/t=8$ and near $1/8$ doping, the CP error in the energy \cite{Hao-Shi-PRB-2013,Hubbard-benchmark} using 
a $|\psi_T\rangle$ from unrestricted Hartree-Fock is less than the Trotter error from a time-step choice of $\tau=0.05\,t^{-1}$, which 
is  typically considered a very conservative choice in standard calculations.)
Better choices of $|\psi_T\rangle$ can reduce the systematic bias. 
In our implementation, 
the trial wave function $|\psi_T\rangle$  is in the form of 
a general Slater determinant: 
\begin{equation}
\label{psiT}
\begin{split}
& | \psi_T \rangle = \prod_{n=1}^{N} \hat{\phi}^{\dagger}_n \, |0 \rangle \\
& \hat{\phi}^{\dagger}_n = \sum_{i=1}^{M} \sum_{\sigma=\uparrow,\downarrow} \sum_{\alpha=d,p_x,p_y} u_n(i,\alpha,\sigma)
\, \hat{\alpha}^{\dagger}_{i,\sigma}
\end{split}
\end{equation}
where the notations follow Eq.~\eqref{3bands:ham}, with the operator $\hat{\alpha}^{\dagger}_{i,\sigma}$
creating a hole of spin $\sigma$ in the $\alpha$-band in the unit cell $i$.

The spin-orbitals $u_n(i,\alpha,\sigma)$ in Eq.~\eqref{psiT}
are constructed within a self-consistent scheme which was introduced in [\onlinecite{Mingpu-sc-PRB}]. 
In the first step a GHF 
calculation is performed 
where the wave function 
\eqref{psiT} is obtained by minimizing the energy $\langle \Psi \, | \, \hat{H} \, | \, \Psi \rangle$ within the manifold of $N$-particles
Slater determinants, using the true Hamiltonian in Eq.~\eqref{3bands:ham}.
For the GHF procedure we do not assume any particular form for the order parameter, 
and we use a combination of randomization and annealing to help find the global minimum \cite{Chiciak_GHF}. 
In the following steps, we use the results of CP-AFQMC 
simulations 
to correct the trial wave function internally \cite{Mingpu-sc-PRB} in which the output of a CP-AFQMC calculation relying on a given  
$|\psi_T\rangle$ is given as feedback in generating a new trial in the GHF framework, but using effective Hamiltonians for $ \hat{H}$.

In practice, the new wave function is found by diagonalizing an effective one-body hamiltonian, 
like in the original GHF procedure, but with effective parameters that are chosen so as 
to minimize the discrepancy between the variational and the CP-AFQMC estimations of the one-body density matrix. 
Then, a new CP-AFQMC calculation relying on the updated $|\psi_T\rangle$ is performed and 
the procedure is continued until convergence is reached. 
This interface between sophisticated mean-field and correlated CP-AFQMC makes our ``adaptive'' 
algorithm able to ``learn'' the best trial wave function to feed the final CP-AFQMC simulation. 

As a further check of the reliability of the approach, 
we systematically explore the robustness of the self-consistency loops against the choice of the initial condition, 
that is the wave-function used in the first iteration.
Although the GHF solution is a natural starting point, we explored 
starting from the non-interacting ground-state, 
as well as from mean-field wave functions displaying other possible orders such as spin density waves, spirals, domain walls. 
As seen in Fig.~\ref{fig:convergence}, the self-consistency loops converge to the same spin order, even starting from an initial state of the 
GHF form made up of random orbitals. 
This is a very strong indication that our calculations 
minimize the bias arising from the constraint to control 
the sign problem, 
and provides another stringent check on the robustness and accuracy of the many-body results.


\subsection{Extrapolation to Thermodynamic Limit}
\label{ssec:TDL}
\begin{figure}[ptb]
\includegraphics[width=\columnwidth, angle=0]{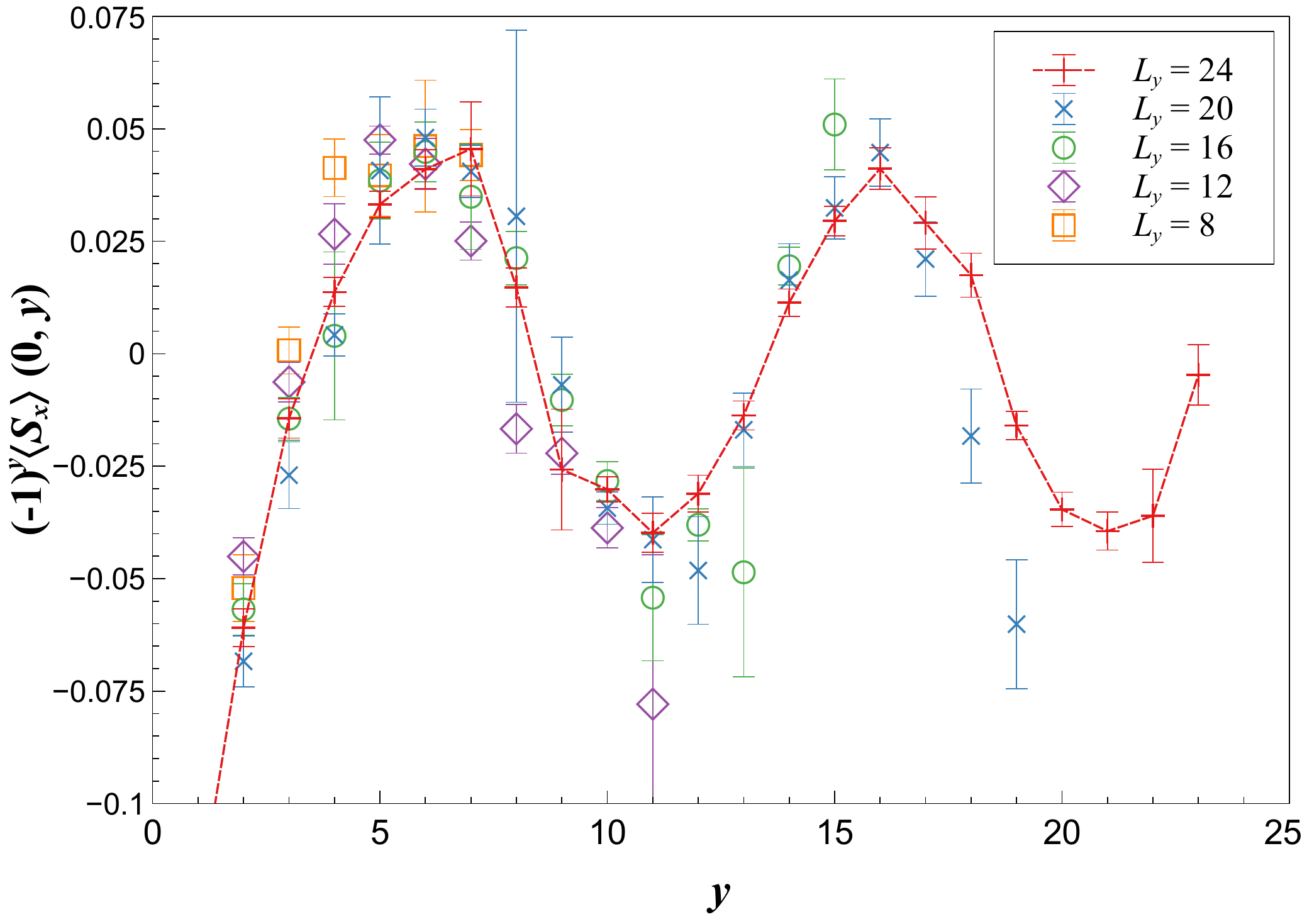}
\caption{
 (Color online)  Plot of 
 the {\it staggered\/} spin vector, $(-1)^y S_x$, for $\Delta = 4.4$ along
 the line cut at $x = 0$ for a group of $6 \times L_y$ systems. 
 A pinning field is applied at $y=0$.
 The spin across the $x$-direction is AFM.
 The spin order converges as $L_y \to \infty$.
 }
\label{fig:thermo}
\end{figure}

Our AFQMC calculations treat large supercells containing up to $\sim 500$ atoms, which makes it possible to capture long wavelength collective modes. 
In addition, we perform various 
tests to help 
extrapolate our results to the bulk limit.
Our calculations in periodic supercells with $L_x \times L_y$ show that the AFQMC solution
favors orders along the $x$ -or $y$-direction, i.e., along 
the lines connecting the $d$ orbitals with the nearest neighbor $p$ orbitals. 
Unlike in the mean-field solutions \cite{Chiciak_GHF}, 
we find no evidence  at the many-body level of a tendency to form long-range orders in the diagonal direction.
On the other hand, our results from varying lattice sizes and aspect ratios 
clearly indicate that large lattices are needed to accommodate the order while minimizing frustration. 

The systematic analysis and experimentation (see for example the results below in Table~\ref{table:properties}) led us to focus on 
studying rectangular $6 \times L_y$ and $8 \times L_y$ systems.
We use periodic boundary
condition (PBC) along the $x$-direction and open boundaries along the $y$-direction, giving the system the topology of a cylinder,  
in such a way to accommodate commensurate spin and charge orders along the $y$ direction. We have also carried out calculations with 
PBCs along both directions (still applying pinning field) to verify the consistency of our results.
The cylinder systems and the pinning field break translational symmetry along $y$ and $C_4$ symmetry, which makes it compatible to use the self-consistent procedure discussed in Sec.~\ref{ssec:self-consistent}.
%
 Figure~\ref{fig:thermo} shows a validation  versus $L_y$, to establish the spin order as $L_y \to \infty$.
We see that,  within statistical error, the spin order is already converged at $L_y = 16$.
Comparing $6 \times L_y$ calculations with $8 \times L_y$ (and wider systems when there is any indication of numerical difference or as spot checks), we validate that 
the dependence on $L_x$, when $L_y$ is large enough, is negligible. 


The external pinning field in Eq.~(\ref{eq:pinning}) 
plays the role of a surface term. For a fixed value of $L_x$, both the energies and the densities show finite-size effects consistent with a linear dependence on $1/L_y$, which allows us to 
extrapolate to the  limit $L_y \to +\infty$. 
This provides results free of the finite-size effects arising from the pinning field and the open boundary condition in the $y$-direction.
We then analyze the effect of increasing $L_x$, the dimension in the periodic direction,  and show that $8\times \infty$ results show 
negligible finite-size errors from the bulk limit. More details are provided in the next section.

\section{Results}
\label{sec:results}

\begin{table*}[t!]
\begin{center}
\caption{
 Table of measured ground-state properties at doping $h=1/8$, for different supercell sizes $M = L_x \times L_y$, at two different values of charge transfer energy $\Delta$.
 All systems have PBC in both directions and have
 a pinning field applied in one row along the short direction with $h_{\rm pinn}=0.05$. 
The quantities are energy per site, 
 $d$- and $p$- (sum of $p_x$ and $p_y$) occupancies, percentage of the doped holes in the Cu $d$ band, 
 expectation values of the hopping matrix elements (kinetic energy components), and the interaction energy.
 }
 \resizebox{\textwidth}{!}{%
\begin{tabular}{c | c | c c c c c c c c c}
\hline\hline
\begin{tabular}{@{}c@{}} $\Delta$          \\ $(eV)$       \end{tabular} &
\begin{tabular}{@{}c@{}} $L_x\times L_y$   \\ $ $          \end{tabular} &
\begin{tabular}{@{}c@{}} $E_{tot}/M$       \\ $(eV)$       \end{tabular} &
\begin{tabular}{@{}c@{}} $n_d$             \\ $ $          \end{tabular} &
\begin{tabular}{@{}c@{}} $n_p$             \\ $ $          \end{tabular} &
\begin{tabular}{@{}c@{}} $\delta_{n_d}$    \\ $(\%)$       \end{tabular} &
\begin{tabular}{@{}c@{}} $T_{dd}$          \\ $(eV)$       \end{tabular} &
\begin{tabular}{@{}c@{}} $T_{pd}$          \\ $(eV)$       \end{tabular} &
\begin{tabular}{@{}c@{}} $T_{pp}$          \\ $(eV)$       \end{tabular} &
\begin{tabular}{@{}c@{}} $E_{int}/M$       \\ $(eV)$       \end{tabular} \\
\hline
$4.4$         & $6\times 8$     & $-10.3389(1)$  & $0.764(1)$  & $0.361(1)$  & $31.7(1)$ &  $0.0538(3)$    & $0.1345(3)$    & $0.0584(3)$    & $0.2006(1)$    \\
              & $6\times 12$    & $-10.3578(1)$  & $0.761(1)$  & $0.364(1)$  & $31.8(1)$ & $0.0539(3)$    & $0.1347(3)$    & $0.0584(3)$    & $0.2023(1)$   \\
              & $6\times 16$    & $-10.3679(1)$  & $0.759(1)$  & $0.366(1)$  & $31.3(1)$ & $0.0543(3)$    & $0.1345(3)$    & $0.0589(3)$    & $0.2036(1)$    \\
              & $6\times 20$    & $-10.3724(1)$  & $0.758(1)$  & $0.367(1)$  & $31.5(1)$ &  $0.0546(3)$    & $0.1346(3)$    & $0.0587(3)$    & $0.2038(1)$    \\
              & $6\times 24$    & $-10.3762(1)$  & $0.758(1)$  & $0.367(1)$  & $31.8(1)$ &  $0.0541(3)$    & $0.1347(3)$    & $0.0587(3)$    & $0.2040(1)$    \\
              & $6\times \infty$& $-10.395(1)$  & $0.754(1)$  & $0.371(1)$  & $31.6(1)$ &                 &                &                & $0.206(1)$    \\
              & $8\times 12$    & $-10.3572(1)$  & $0.761(1)$  & $0.364(1)$  & $31.9(1)$ & $0.0539(3)$    & $0.1339(3)$    & $0.0587(3)$    & $0.2025(1)$    \\
              & $8\times 14$    & $-10.3627(1)$  & $0.760(1)$  & $0.365(1)$  & $32.0(1)$ & $0.0541(3)$    & $0.1351(3)$    & $0.0585(3)$    & $0.2026(1)$    \\
              & $8\times 16$    & $-10.3658(1)$  & $0.760(1)$  & $0.365(1)$  & $32.3(1)$ &  $0.0533(3)$    & $0.1348(3)$    & $0.0588(3)$    & $0.2027(1)$    \\
              & $8\times 18$    & $-10.3692(1)$  & $0.759(1)$  & $0.366(1)$  & $31.8(1)$ & $0.0539(3)$    & $0.1343(3)$    & $0.0589(3)$    & $0.2037(1)$    \\
              & $8\times 20$    & $-10.3718(1)$  & $0.758(1)$  & $0.366(1)$  & $31.8(1)$ &  $0.0542(3)$    & $0.1339(3)$    & $0.0588(3)$    & $0.2037(1)$    \\
              & $8\times \infty$& $-10.393(1)$  & $0.754(1)$  & $0.371(1)$  & $31.9(1)$ &                 &                &                & $0.206(1)$    \\
\hline
$2.5$         & $6\times 8$    & $-9.0480(1)$  & $0.594(1)$ & $0.531(1)$ & $36.9(1)$  &  $0.0644(1)$    & $0.1512(1)$    & $0.0847(1)$    & $0.1930(1)$    \\
              & $6\times 12$   & $-9.0737(1)$  & $0.591(1)$ & $0.534(1)$ & $36.2(1)$  &  $0.0641(1)$    & $0.1505(1)$    & $0.0844(1)$    & $0.1941(1)$    \\
              & $6\times 16$   & $-9.0877(1)$  & $0.589(1)$ & $0.536(1)$ & $35.7(1)$  & $0.0639(1)$    & $0.1503(1)$    & $0.0848(1)$    & $0.1950(1)$    \\
              & $6\times 20$   & $-9.0942(1)$  & $0.588(1)$ & $0.537(1)$ & $36.1(1)$  &  $0.0636(1)$    & $0.1498(1)$    & $0.0846(1)$    & $0.1949(1)$    \\
              & $6\times 24$   & $-9.0989(1)$  & $0.588(1)$ & $0.537(1)$ & $36.0(1)$  &  $0.0637(1)$    & $0.1501(1)$    & $0.0843(1)$    & $0.1951(1)$    \\
              & $6\times \infty$& $-9.125(1)$  & $0.585(1)$  & $0.540(1)$  & $35.4(1)$ &               &                &                & $0.196(1)$    \\
              & $8\times 12$   & $-9.0719(1)$  & $0.592(1)$ & $0.533(1)$ & $36.6(1)$  &  $0.0630(1)$    & $0.1498(1)$    & $0.0846(1)$    & $0.1937(1)$    \\
              & $8\times 14$   & $-9.0794(1)$  & $0.592(1)$ & $0.533(1)$ & $36.9(1)$  & $0.0630(1)$    & $0.1501(1)$    & $0.0841(1)$    & $0.1935(1)$    \\
              & $8\times 16$   & $-9.0849(1)$  & $0.591(1)$ & $0.533(1)$ & $37.1(1)$  &  $0.0629(1)$    & $0.1500(1)$    & $0.0841(1)$    & $0.1934(1)$    \\
              & $8\times 18$   & $-9.0886(1)$  & $0.590(1)$ & $0.535(1)$ & $36.7(1)$  & $0.0628(1)$    & $0.1498(1)$    & $0.0843(1)$    & $0.1939(1)$    \\
              & $8\times 20$   & $-9.0917(1)$  & $0.590(1)$ & $0.535(1)$ & $36.7(1)$  & $0.0624(1)$    & $0.1497(1)$    & $0.0840(1)$    & $0.1939(1)$    \\
              & $8\times \infty$& $-9.122(1)$  & $0.587(1)$  & $0.538(1)$  & $37.0(1)$ &               &                &                & $0.194(1)$    \\
\hline\hline
\end{tabular}%
}
\label{table:properties}
\end{center}
\end{table*}

In Table~\ref{table:properties} we list the values of several properties of the systems 
as a function of the size of the system $M = L_x \times L_y$ and of the charge transfer energy, 
$\Delta = \epsilon_d - \epsilon_p$. The detailed data may prove useful for future analysis.  
With the high accuracy of these calculations, the results will also help provide benchmark for future studies.
In addition, the details help illustrate the convergence with respect to system size. 

Results are shown for the total energy per site, 
 the kinetic energies measured by the average nearest neighbor hopping amplitudes, 
 which are the lattice averages of the matrix elements of the one-body density matrix
(per site):
$T_{dd}=\langle \hat{d}^{\dagger}_{i}  \hat{d}^{}_{i+\hat{x}(\hat{y})} \rangle$, 
$T_{pd}=\langle \hat{d}^{\dagger}_{i}  \hat{p}^{}_{i+\hat{x}(\hat{y})/2} \rangle$,
$T_{pp}=\langle \hat{p}^{\dagger}_{i+\hat{y}/2}  \hat{p}^{}_{i+\hat{x}/2} \rangle$, 
and the interaction energy.
Also shown are 
the average density of holes on the $d$ and $p$ orbitals respectively (Eq.~(\ref{eq:def-n})), 
 and
the percent of doped holes on the copper $d$-band, defined as 
$\delta_{n_d} = ({n_d}^h - {n_d}^{0})/h$  where the reference ${n_d}^{0}$ 
is the  average density of holes on the $d$ orbitals at half filling, while ${n_d}^h$ is the value at the current doping, $h$.
The quantity gives an indication of the fraction of the doped holes which go on the $d$ sites.

\COMMENTED{
As mentioned, all of our calculations include an external pinning field of the form in Eq.~(\ref{eq:pinning}) 
applied on the line $y=0$,
playing the role of a surface term, explicitly breaking translational and rotational symmetry. For a fixed value of $L_x$, both the energies and the densities appear to be consistent with a linear dependence on $1/L_y$, which allows us to 
extrapolate to the  limit $L_y \to +\infty$ where finite-size effects arising from the pinning field and the open boundary condition in the $y$-direction vanish.
In the table, we report the results of the linear extrapolation as $L_x \times \infty$. 
}

In Table~\ref{table:x-extrap}, 
we 
further examine the behavior of the total energy as a function of the width $L_x$ for fixed $L_y$.
 The PBC helps to  significantly reduce the finite-size effects from $L_x$, which is confirmed by the results showing up to $L_x=12$.
Changing $L_x$ results in variations which are of order ${\mathcal O}(10^{-3})$\,eV, 
consistent with the fact that the $6\times \infty$ and $8\times \infty$ results in   Table~\ref{table:properties}  are in agreement to within this level. 
Thus we expect that the  $8\times \infty$ results listed in Table~\ref{table:properties}, to within the indicated statistical uncertainties,
 are representative of the bulk limit.

\begin{table}
\begin{center}
\caption{Convergence versus supercell size in the periodic direction. 
  The measured ground-state energy per cell, $E_{tot}/M$,
  is shown at doping $h=1/8$, 
  for different supercell sizes $M = L_x \times 16$, at two different values of charge transfer energy $\Delta$.
  All supercells use the same systematic parameters as in Table~\ref{table:properties}.
  }
\begin{tabular}{c | c | c }
\hline\hline
$L_x\times L_y$   & $\Delta = 4.4$ & $\Delta = 2.5$ \\
\hline
$6\times 16$      & $-10.3679(1)$ & $-9.0877(1)$ \\
$8\times 16$      & $-10.3658(1)$ & $-9.0849(1)$ \\
$10\times 16$     & $-10.3672(1)$ & $-9.0860(1)$ \\
$12\times 16$     & $-10.3682(1)$ & $-9.0856(1)$ \\
\hline \hline
\end{tabular}
\label{table:x-extrap}
\end{center}
\end{table}

 Table~\ref{table:properties} provides a first answer to the question: 
 where do the doped holes go, as we move from the parent compound to the underdoped systems?
Expectedly, as $\Delta$ 
is increased, 
the Cu 
$d$-orbital occupation increases both in the half--filled and the doped systems.
The fraction 
of doped holes on the Cu $d$-bands remains smaller than $50\%$ for both values of $\Delta$. 
This means that as holes are doped, 
significantly more 
choose to occupy the $p$-bands over the Cu $d$-bands, 
giving a roughly equal distribution of the excess holes on the $d$ and the two $p$ sites.
Interestingly, while the occupancy of $d$-bands is considerably higher at larger $\Delta$, 
 the percentage of the doped holes on the $d$-bands is slightly lower.

Comparing to the experimental results of Jurkutat et.~al.\cite{Jurkutat_NMR}, 
 our computed orbital occupancies 
 for both the half-filling and 1/8-doped systems are very close to 
 the experimentally measured values in the Y-based cuprate family:
 $n_d\approx 0.75$ and $n_p\approx 0.4$ at $h\approx 0.15$.
At $\Delta = 2.5$, 
our 
computed occupancies are remarkably close to those measured in
the Hg-, Bi-, and TI-based cuprate families, 
with $n_d\approx 0.59$ and $n_p\approx 0.54$  at $h\approx 0.13$.
Furthermore,  the computed $\Delta$-dependence of the percentage of the doped holes occupying the $d$-bands is 
consistent with experiment. 
The results in  [\onlinecite{Jurkutat_NMR}] suggest that
 the 
 distribution of excess holes varies significantly across the different families, and the percentage of holes occupying the $d$-orbitals is significantly larger in the Hg-, Bi-, and TI-based families compared to the  Y-based  family, again consistent with our results. 
 These observations indicate that the three-band Hubbard model indeed captures additional materials specificity which is lacking in the one-band Hubbard model.
 Additionally, the orbital occupancy agreement with experiment suggests empirically that the particular values of $\Delta$ (and other Hamiltonian parameters) 
 are likely good choices to model the two groups of cuprate families.

\subsection{Spin and Charge Orders}
\label{ssec:spin-charge}

\begin{figure}[ptb]
\includegraphics[width=6.0cm, angle=0]{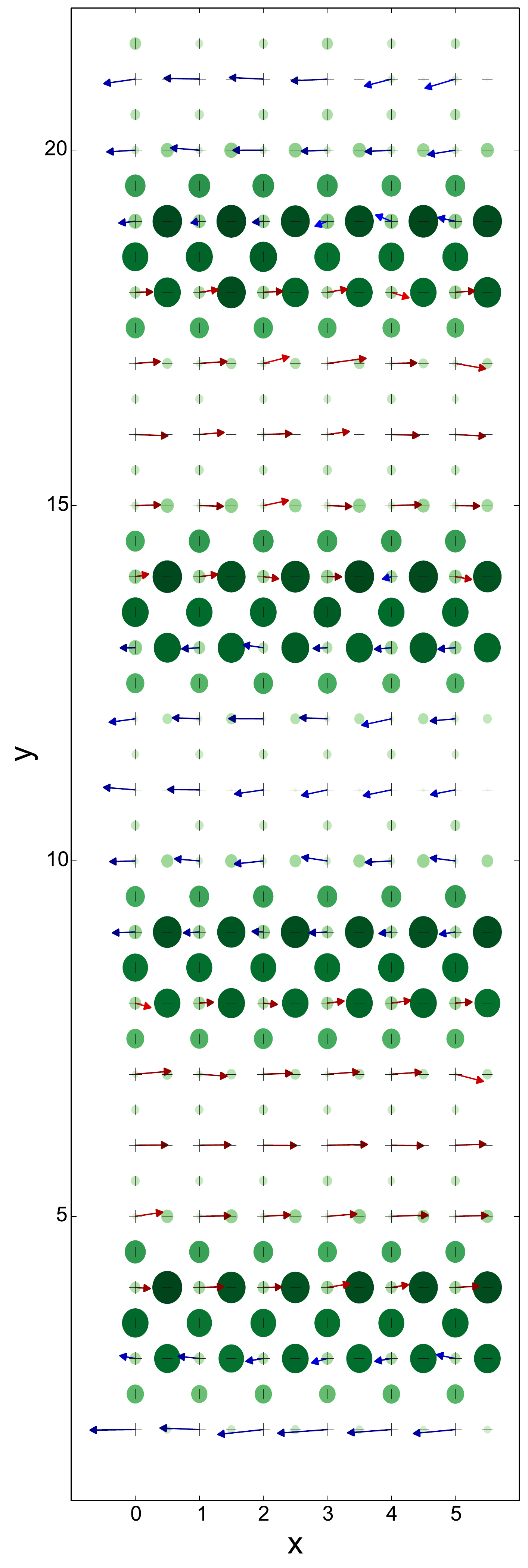}
\caption{
 (Color online) 2-D plot of the staggered spin vector, $(-1)^{x+y} \langle\hat{\mathbf S}({\mathbf r}) \rangle$, and charge density,
$ \langle\hat{n}({\mathbf r}) \rangle$, for $\Delta = 4.4$ and $h=1/8$.
 The total staggered spins (arrows) are plotted as a projection in the $x$-$z$ plane. The color of the arrow represents the angle between
 the spin on that site and an arbitrary reference spin. It can be thought of as a spin correlation and it runs from $(0, \pi)$.
 The spin on the O $p$-orbitals is negligible and omitted from the plot.
 The size of the circle is proportional to the density, with an overall background subtracted away.
 We neglect the first and last two rows to avoid the open boundaries and pinning field.
 }
\label{fig:4.4angle}
\end{figure}

\begin{figure}[ptb]
\includegraphics[width=\columnwidth, angle=0]{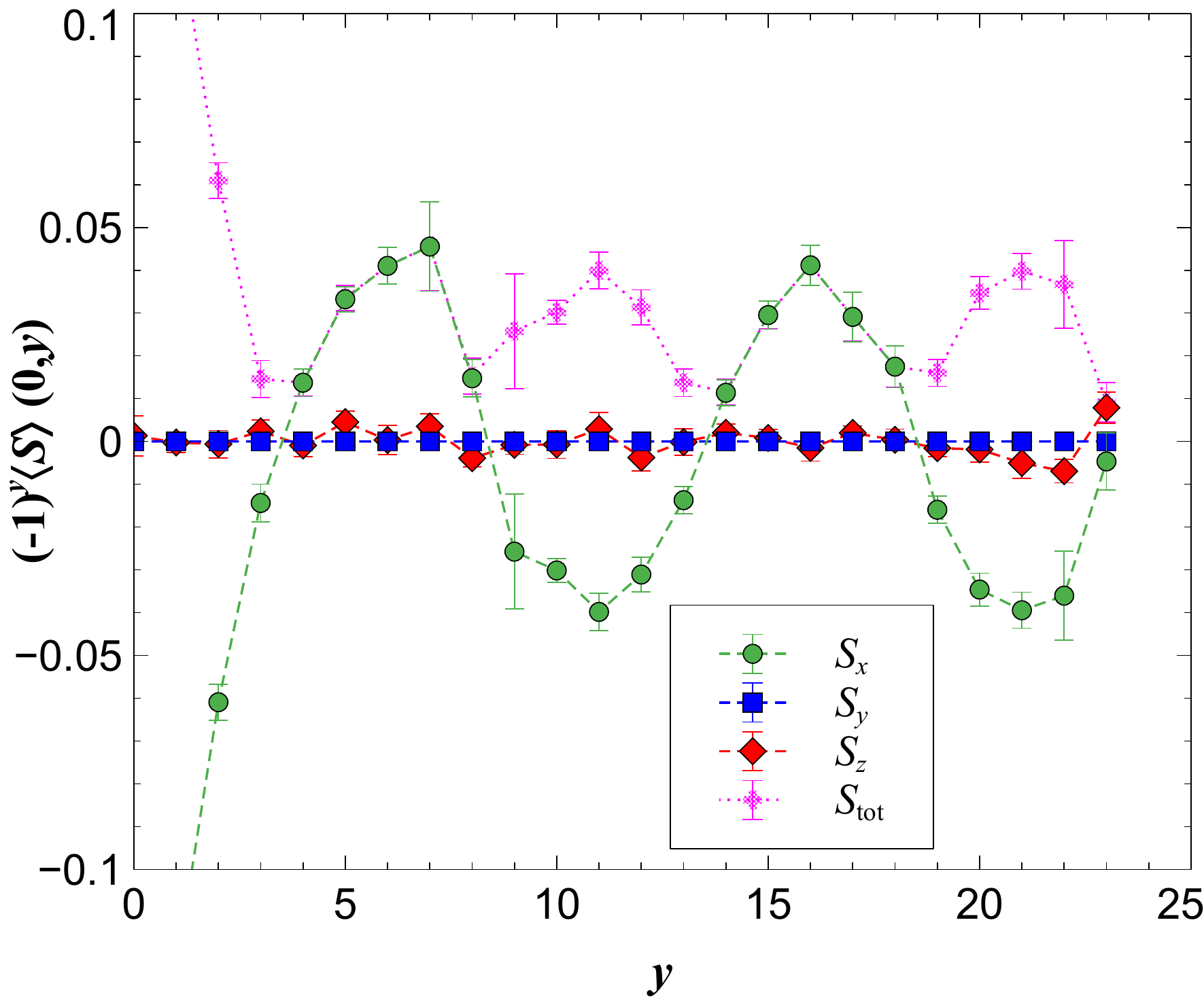}
\caption{
 (Color online) Plot of the components of the staggered spin vector along
 the line cut at $x = 0$, for the system in Fig.~\ref{fig:4.4angle}. The spin across the x--direction is AFM.
 The majority of the spin vector lies in the $S_x$-direction, the same as the pinning field.
 A stripe phase with AFM domains is seen.
 }
\label{fig:4.4spin}
\end{figure}

\begin{figure}[ptb]
\includegraphics[width=\columnwidth, angle=0]{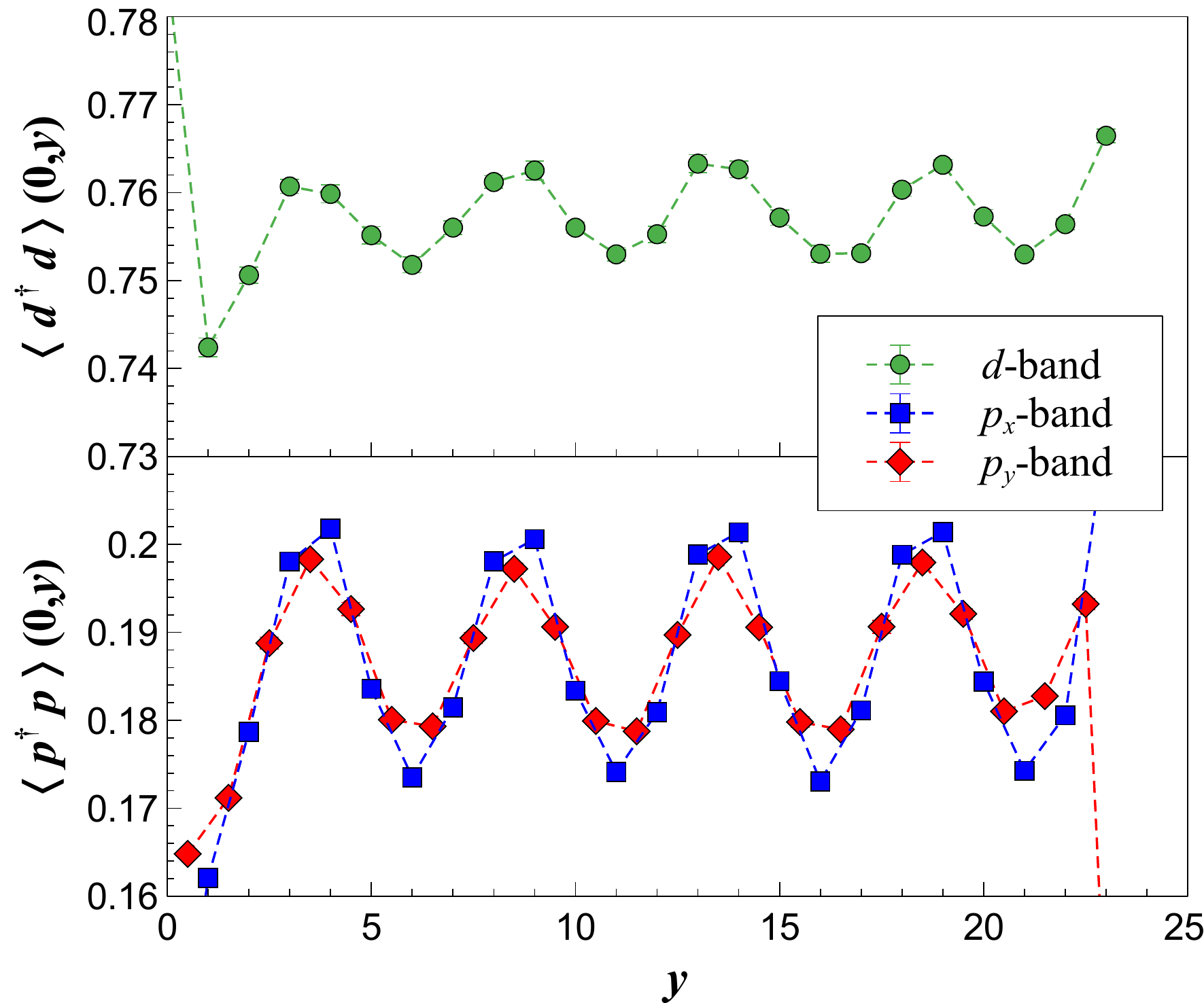}
\caption{
 (Color online) Plot of the occupations on different sites 
along the line
 cut at $x = 0$, for the system in Fig.~\ref{fig:4.4angle}. The Cu $d$-band occupation is shown in the top panel, and 
 the O $p_x$ and $p_y$-bands are plotted in the bottom. The hole density wave is correlated with the spin order in Fig.~\ref{fig:4.4spin},
 with higher density at the domain boundaries. A small asymmetry is seen between $p_x$ and $p_y$ sites.
 }
\label{fig:4.4dens}
\end{figure}

We find that the spin--orders 
in the Emery model tend to be very 
subtle, with multiple viable orders competing at tiny energy scales. 
This results in a high sensitivity of the spin-order with respect to the details of the trial wave function guiding 
the CP-AFQMC procedure and with respect to the size of the system. 
It was necessary to perform systematic crosschecks by initializing 
the self-consistent loop described in Sec.~\ref{ssec:self-consistent}
in several different ways: 
diagonal magnetic domain walls, spin-density waves (SDW), spiral orders, 
and homogeneous phases were used as initial trial wave functions. 
After several iterations, consistency is reached in many cases, allowing us to draw conclusions 
about the spin order 
in the ground state of the model as a function of the charge--transfer energy. 
We will highlight cases where different candidate spin orders are especially close and the balance is especially delicate, as indicated by 
the competition persisting with the self-consistency, and by closeness of their energies.
The charge--order, on the other hand, appears to be very 
robust.  
Negligible effects are seen of 
the choice of the trial wave-function and of the system size 
on 
the density of holes on $d$ and $p$ orbitals. 

At the higher value of the charge-transfer energy, $\Delta = 4.4$, a stripe--like phase appears. 
The spin and charge orders are illustrated in 
Fig.~\ref{fig:4.4angle}.
The spin density on the  $p$ orbitals turns out to be negligible, so we only show the spin order on the Cu $d$ orbitals. 
Figure~\ref{fig:4.4spin} shows the spin order in more details, where spatially modulated spin densities along the $y$-direction are seen. 
The majority of the spin vector lies in the $S_x$-direction, the same as the pinning field.
Figure~\ref{fig:4.4dens} shows the charge occupations on the Cu $d$- and the O $p_x$- and $p_y$-sites, along the same line cut
as the spin density above.
%
From these figures 
we can visualize a regular distribution of AFM domains, 
separated by regions of high holes density, in particular on the $p$ orbitals, 
where the AFM order 
reverses direction. 
The ``node'' where the reversal occurs falls between two  Cu sites, creating a ``domain wall'' between two AFM domains with 
 two  adjacent rows of aligned spins on the Cu $d$-orbitals.
 The wavelength of the spin order on the $d$ orbitals is around 10 Cu sites, while hole densities show a corresponding oscillation with half the 
period and higher density tending towards the domain boundary of the spin order. 
These characters are similar to the behavior of stripe orders seen in the one-band Hubbard model.

\begin{figure}[ptb]
\includegraphics[width=6.0cm, angle=0]{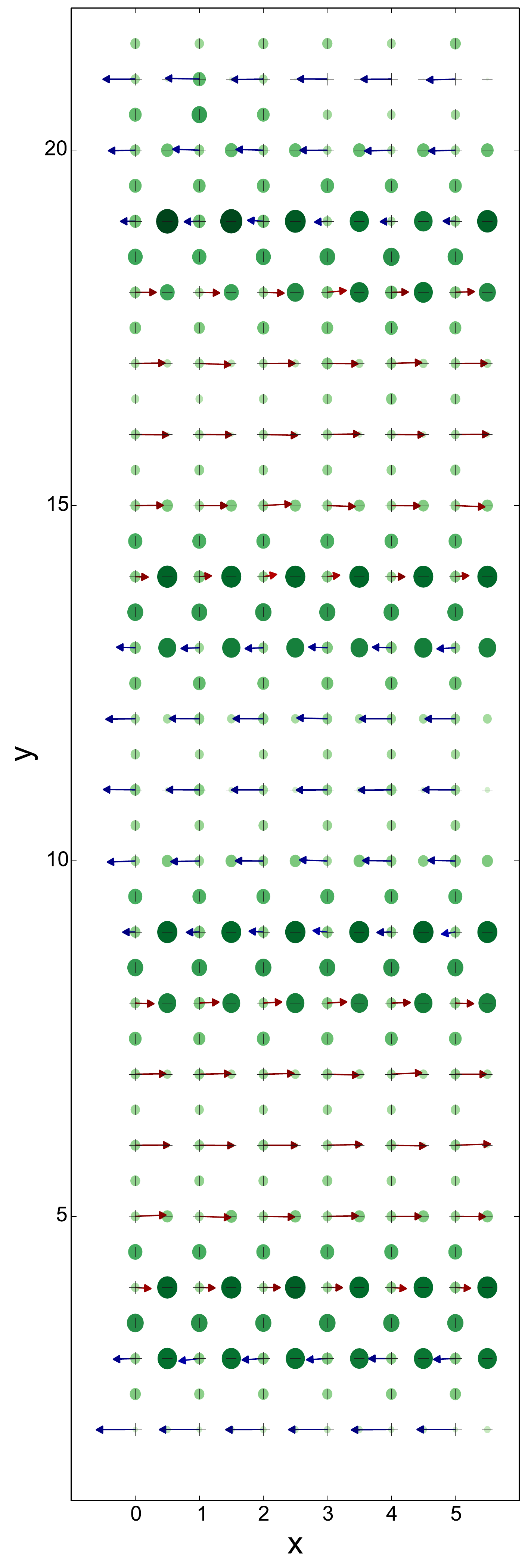}
\caption{
 (Color online) 2-D plot of the staggered spin vector and hole density, similar to Fig.~\ref{fig:4.4angle}, but
  for $\Delta = 2.5$.
 }
\label{fig:2.5angle}
\end{figure}

\begin{figure}[ptb]
\includegraphics[width=\columnwidth, angle=0]{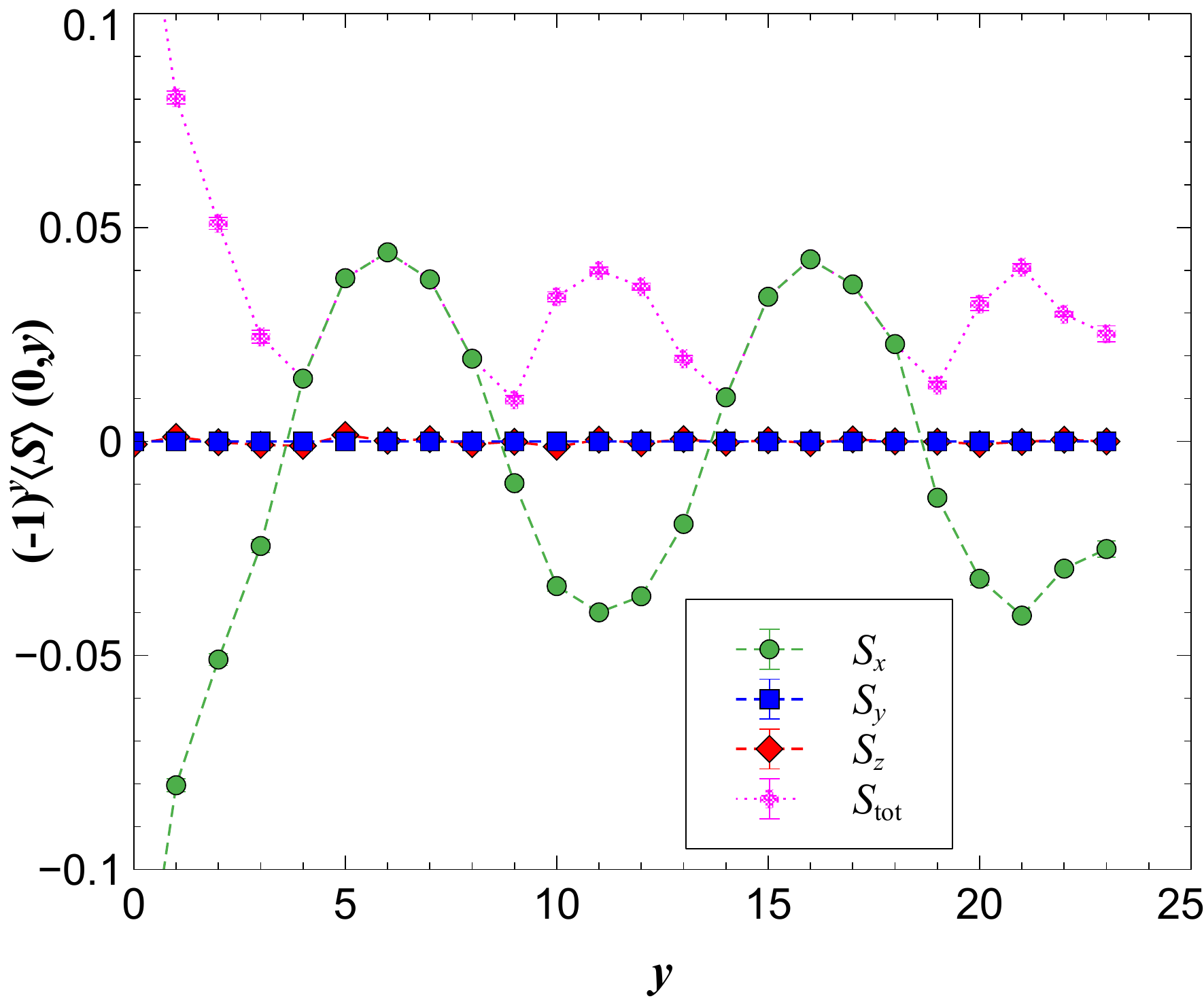}
\caption{
 (Color online) Plot of the components of the staggered spin vector along
 the line cut at $x = 0$, for the system in Fig.~\ref{fig:2.5angle}. The spin across the x--direction is AFM.
 The majority of the spin vector lies in the $S_x$-direction, the same as the pinning field.
 A smooth AFM spin-density wave is seen.
 }
\label{fig:2.5spin}
\end{figure}

\begin{figure}[ptb]
\includegraphics[width=\columnwidth, angle=0]{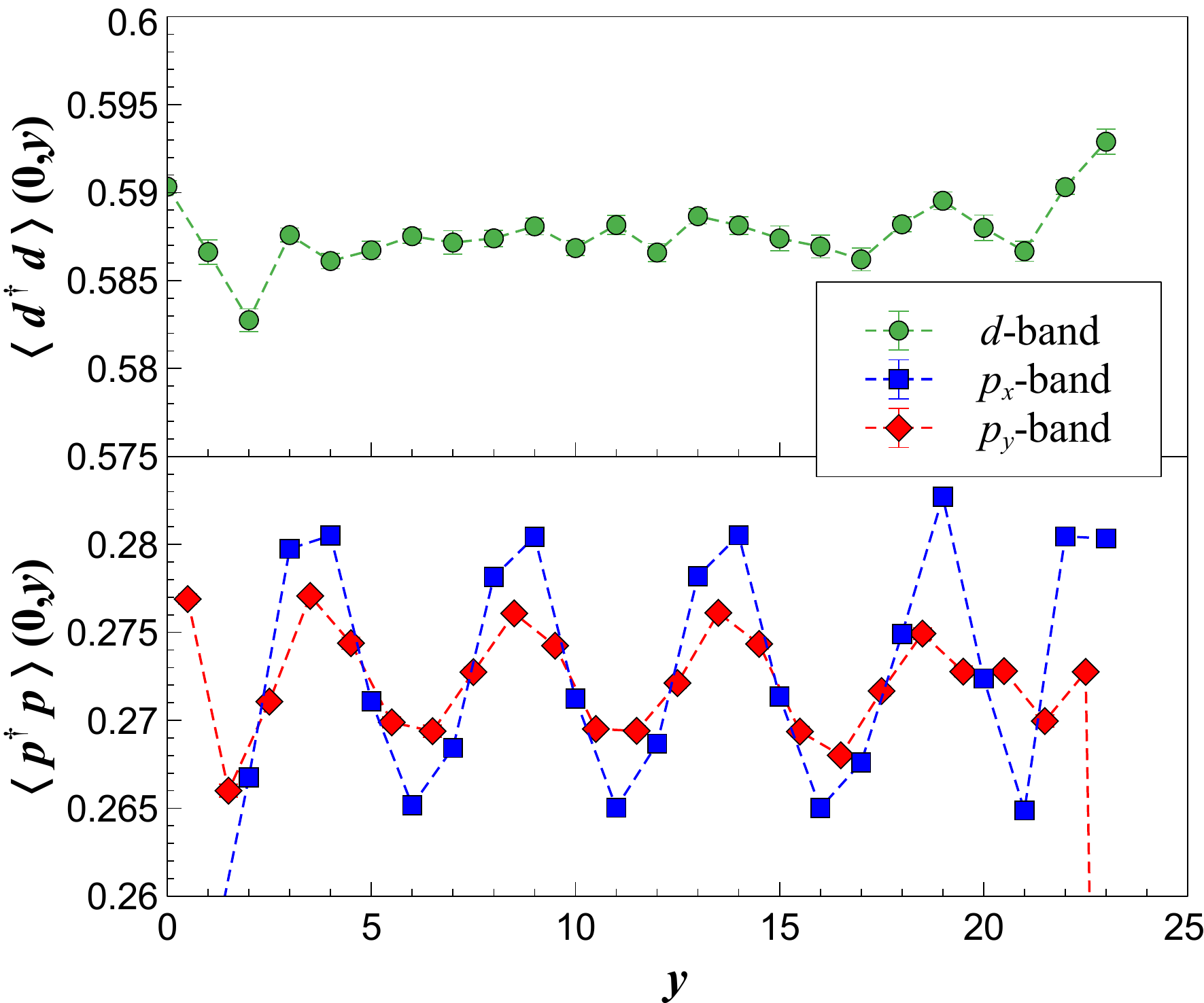}
\caption{
 (Color online) 
 Plot of the occupations on different sites 
along the line
 cut at $x = 0$, for the system in Fig.~\ref{fig:2.5angle}. The Cu $d$-band occupation is shown in the top panel, and 
 the O $p_x$ and $p_y$-bands are plotted in the bottom.  
 Densities on the $d$ sites show little fluctuation, while occupations on the $p$ sites are correlated with the spin density in 
 Fig.~\ref{fig:2.5spin}, with $p_x$ sites showing a much larger response.  }
\label{fig:2.5dens}
\end{figure}

The results at lower $\Delta = 2.5$ are shown in Figs.~\ref{fig:2.5angle}, \ref{fig:2.5spin} and \ref{fig:2.5dens}.
The spin order appears to be substantially smoother than at $\Delta = 4.4$.
We interpret this as a signature of a 
shift toward a SDW phase, in contrast with the 
situation at $\Delta = 4.4$ which suggests a stripe-like order.
For the charge order, the average Cu $d$-orbital occupation is nearly uniform
and, as expected,  greatly reduced with respect to $\Delta = 4.4$. There are still signs of
a charge density wave on the O $p$-orbitals, although the amplitude is decreased 
by half compared to the charge wave at $\Delta = 4.4$. 
The maxima of the density of holes on the $p$ orbitals correspond to the nodes of the staggered spin density on the $d$ orbitals, 
as happens at the higher $\Delta$. A significant asymmetry is seen in the occupancy of the O $p_x$ and $p_y$ sites, indicative of 
a strong nematic response to the SDW.

\begin{figure}[ptb]
\includegraphics[width=\columnwidth, angle=0]{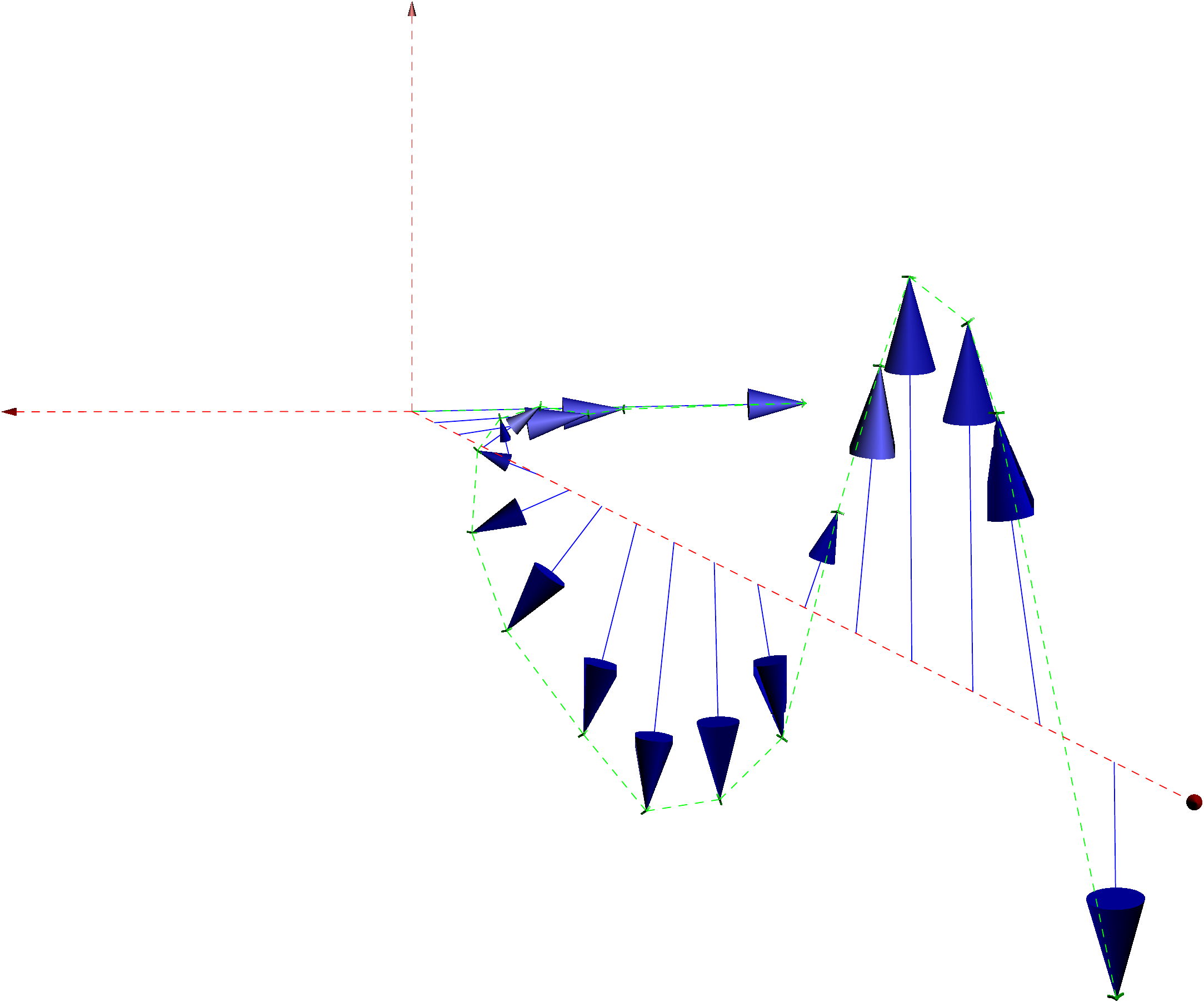}
\caption{
 (Color online) 3D plot of the staggered spiral spin order for an $8 \times 18$ system, at $\Delta = 2.5$ and $h=1/8$.
The staggered spin is shown, projected in three-dimensions along a line-cut at $x = 0$ plotting along the $y$-direction.
 Along the $x$-direction, the order remains perfect AFM.
 }
\label{fig:spiral}
\end{figure}

We find 
that a spiral order, pictured in Fig.~\ref{fig:spiral},
can become stable at $\Delta = 2.5$, and 
is nearly degenerate with respect to the SDW order within our resolution. 
The nature of the spiral order is similar to that seen in the generalized Hartree-Fock solution \cite{Chiciak_GHF}.
The AFQMC self-consistency loops can converge to a spiral state or a SDW depending 
on the starting trial wave function, and the resulting 
energies are extremely close. 
For example,  in an $8\times18$ supercell the energy per site is $-9.0881(1)$ for the SDW state, versus $ -9.0886(1)$ for the spiral state.
The state also depends delicately on the details of the system. 
As in $8\times18$, the $6\times18$ system also shows the spiral state as having slightly lower energy; however, in the $8\times20$ supercell 
the energy ordering is reversed. 
We conclude that there is an extremely subtle cooperation or competition between 
the SDW and spiral phases in this region of the phase diagram. 
This suggests that, in the ground state of the Emery model, when the charge transfer energy is small, 
the spin order appears to be relatively ``soft'', while the charge density 
appears to be more homogeneous 
compared to higher values of $\Delta$.

As mentioned above, our explorations indicate that the
charge and spin orders in the ground state of the Emery model, 
 for  the parameters studied in this work, appear along the $x$- or $y$-direction, i.e., the
direction connecting a Cu site to one of its nearest neighbor O site. 
This 
 led us to focus on elongated geometries of supercells,
in order to accommodate potential collective modes. 
The artificial symmetry-breaking makes it easier to probe the density waves, but more delicate to study nematic orders, 
especially with the necessary reduction in supercell size in QMC compared to mean-field calculations.
In the latter, nematic orders readily appeared for intermediate $\Delta$ values \cite{Chiciak_GHF}.
Intra--unit cell nematic order 
has been observed both in theory \cite{Zegrodnik_nematic, Zegrodnik_VMC} 
and experiment\cite{Comin_nematic}. 
Within our QMC calculations, 
signatures of nematicity are present in 
narrow $4 \times L_y$ systems; 
as $L_x$ is increased, the spatially averaged nematic order $|n_{p_x} - n_{p_y}|$
fades away.
However, locally, on the unit cell, nematic order is present in Fig.~\ref{fig:4.4dens} and is very apparent at lower $\Delta$ in Fig.~\ref{fig:2.5dens}. 
This local nematic order accompanies 
the long-range spin and charge orders, 
which explicitly break the rotational symmetry in the lattice
and in which 
the doped holes tend to organize close to the nodes of the spin density to induce asymmetry.

\subsection{Momentum Distributions}
\label{ssec:momentum}

We also compute 
the momentum distribution of the holes  in the Emery model: 
\begin{equation}
\label{nofk}
n_\sigma (\bm{k}) = \left\langle \hat{d}^{\dagger}_{\bm{k},\sigma} \hat{d}^{}_{\bm{k},\sigma} 
+ \hat{p}^{\dagger}_{x, \, \bm{k},\sigma} \hat{p}^{}_{x,\,\bm{k},\sigma} 
+ \hat{p}^{\dagger}_{y, \, \bm{k},\sigma} \hat{p}^{}_{y,\,\bm{k},\sigma} 
\right\rangle\,,
\end{equation}
where the creation (destruction) operators are the Fourier components of the operators appearing in the Hamiltonian in Eq.~\eqref{3bands:ham}.
Each of the three terms on the right-hand side of Eq.~(\ref{nofk}) gives a \emph{band-resolved contribution}, which we will also examine separately below.
We focus on the stripe phase at $\Delta = 4.4$ and on the spiral phase at $\Delta = 2.5$.

\begin{figure}[ptb]
\includegraphics[width=\columnwidth, angle=0]{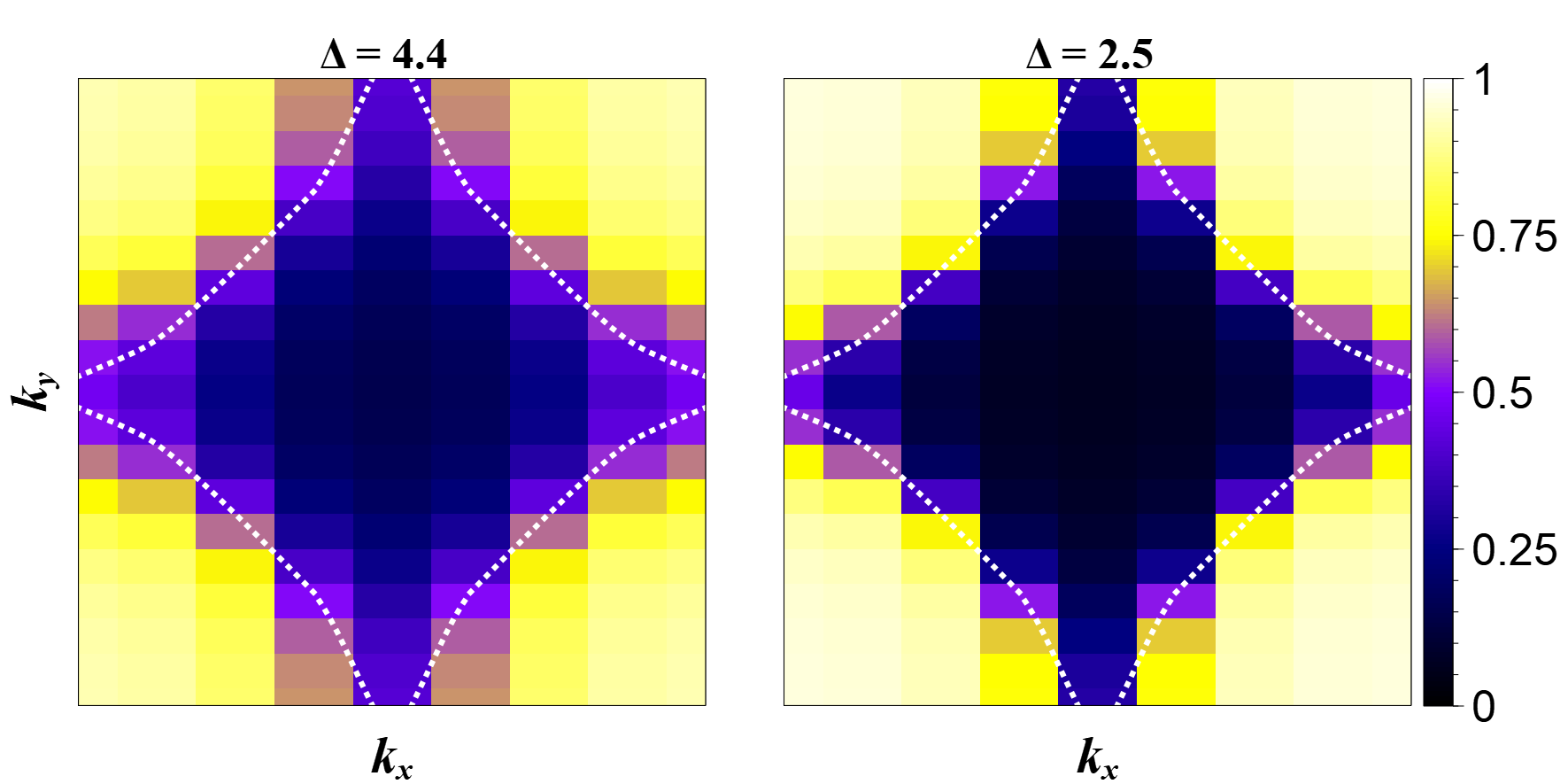}\\
\includegraphics[width=\columnwidth, angle=0]{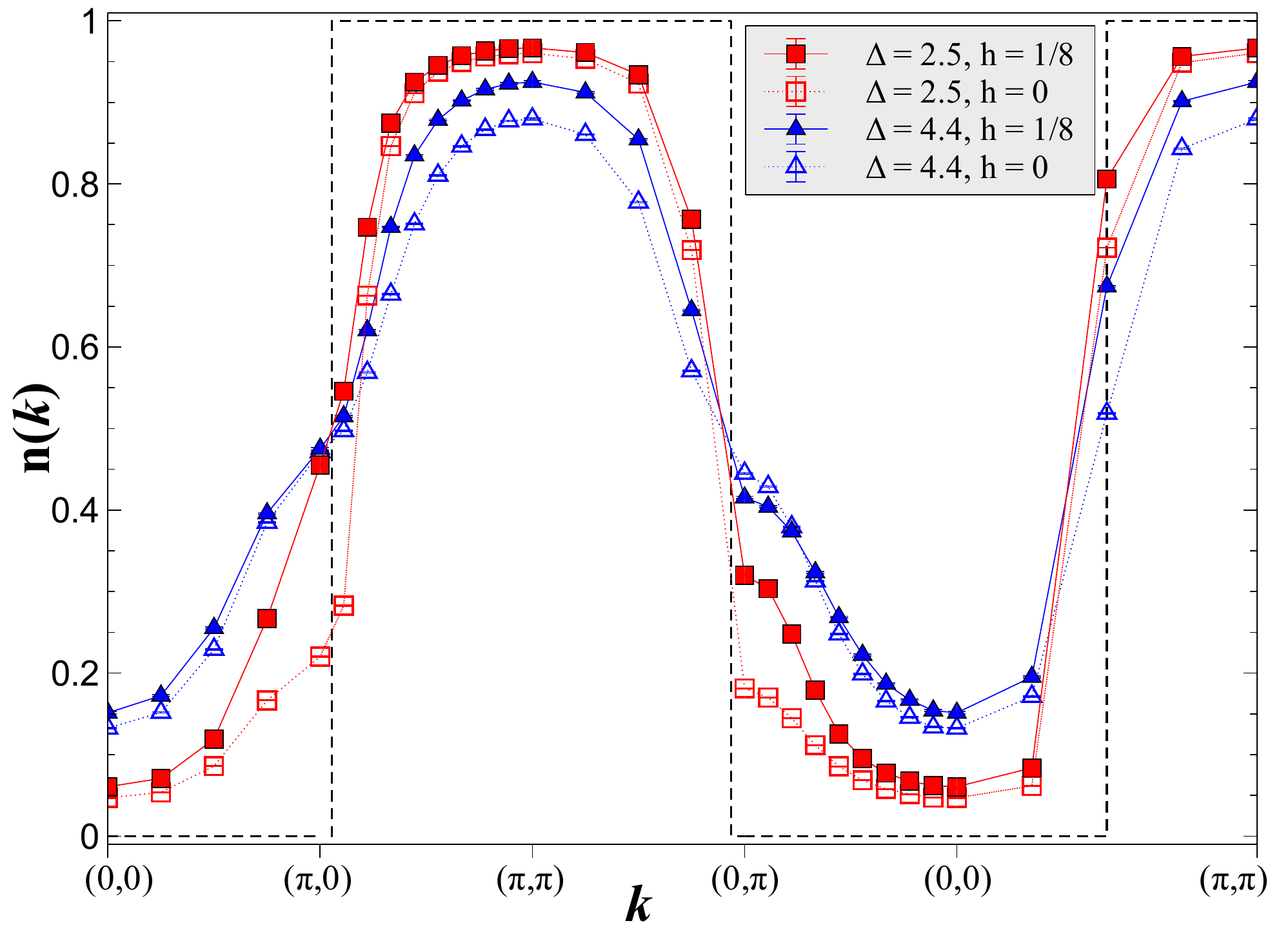}
\caption{
 (Color online) (Top) Total momentum distribution, $n(\bm{k})$
 in the $8\times18$ system at $h=1/8$,  for $\Delta = 4.4$ (left) and  $\Delta = 2.5$ spiral state (right). For reference, the corresponding 
 non-interacting Fermi surface is plotted as a white dashed line.
 (Bottom) 
 $n(\bm{k})$ plotted
 along the path in momentum space $(0,0) \rightarrow (\pi,0) \rightarrow (\pi,\pi) \rightarrow (0,\pi) \rightarrow (0,0) \rightarrow (\pi,\pi)$
 for the same systems in (a), together with their corresponding half-filled systems. 
 For reference, the non-interacting  $n(\bm{k})$ is plotted as the black dashed line.
 }
\label{fig:mom_dist_both}
\end{figure}

\begin{figure}[ptb]
\includegraphics[width=\columnwidth, angle=0]{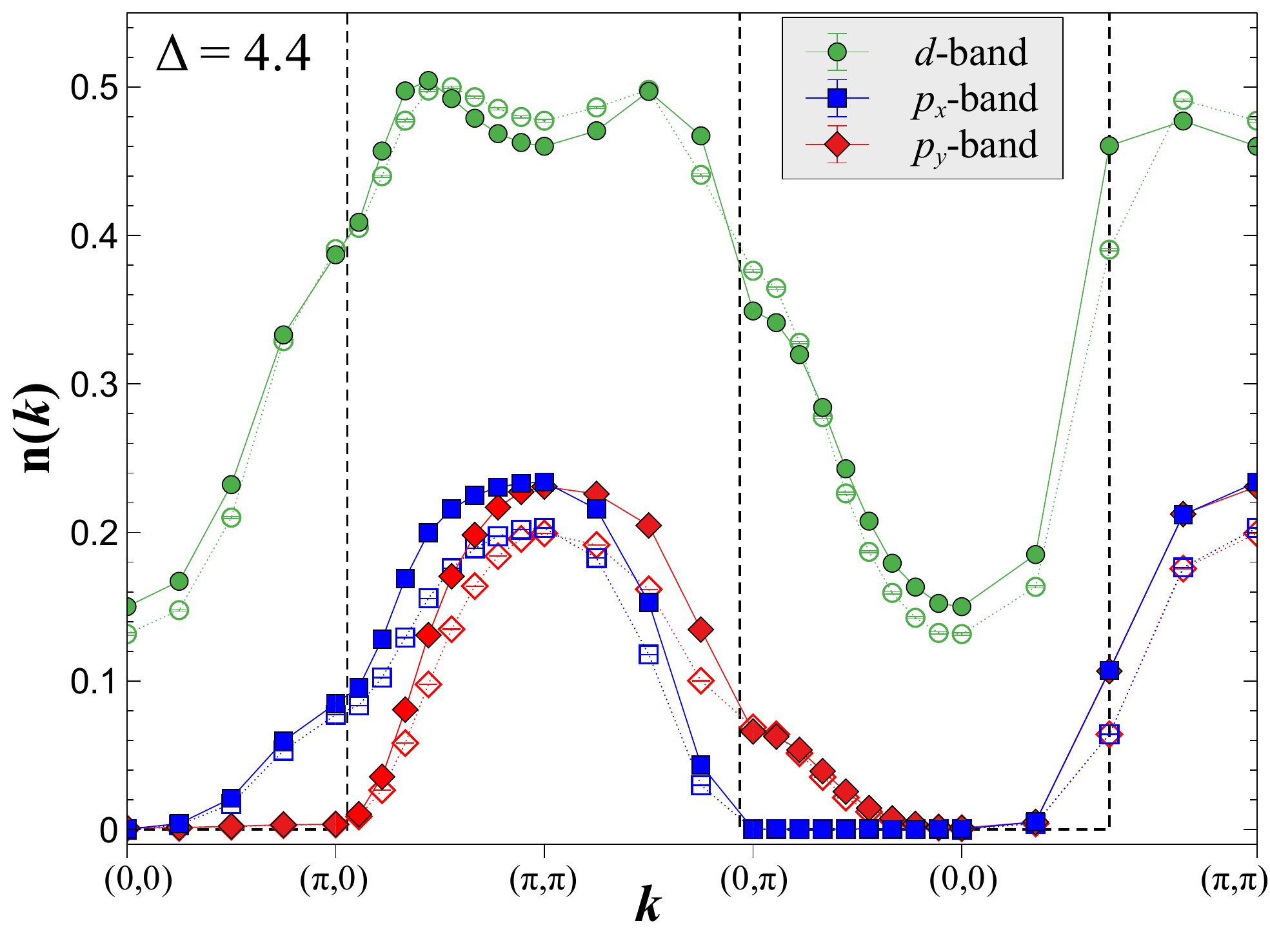}\\
\includegraphics[width=\columnwidth, angle=0]{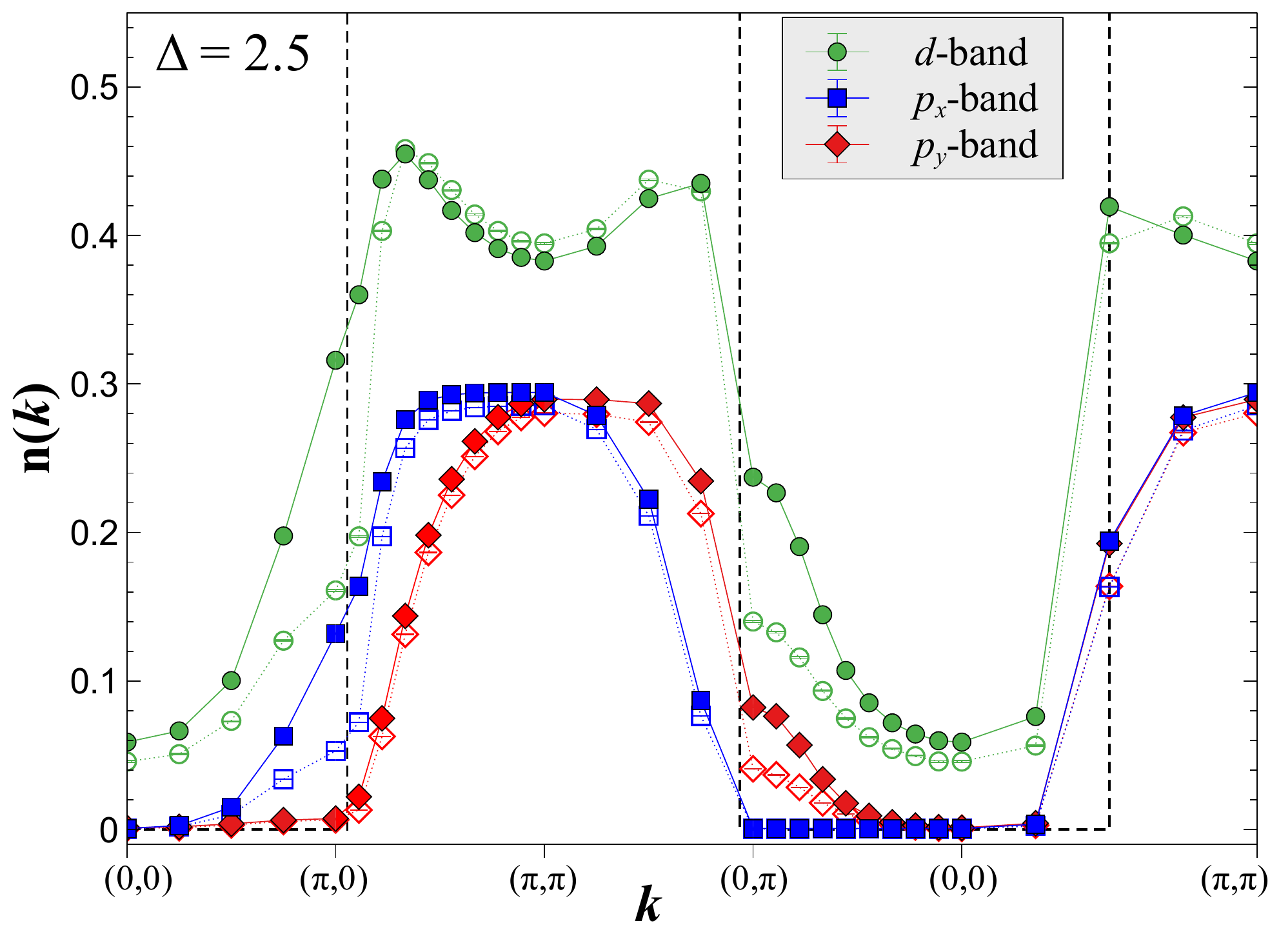}
\caption{
 (Color online) Band resolved 
 momentum distributions, plotted
 along the path 
 $(0,0) \rightarrow$ $(\pi,0) \rightarrow$ $(\pi,\pi) \rightarrow $ $(0,\pi) \rightarrow $ $(0,0) \rightarrow (\pi,\pi)$
 for the same systems as in Fig.~\ref{fig:mom_dist_both}. Filled symbols denote the $h=1/8$  doped systems, while open symbols 
 the corresponding half-filled systems.
 For reference, the non-interacting  $n(\bm{k})$  is plotted as the black dashed line.
 }
\label{fig:mom_dist_pathbr}
\end{figure}

The top panel Fig.~\ref{fig:mom_dist_both} 
shows the total momentum distributions $n(\bm{k}) \equiv \frac{1}{2}(n_{\uparrow}(\bm{k}) + n_{\downarrow}(\bm{k})) $ 
 in an $8 \times 18$ lattice  
for $\Delta = 4.4$ and $\Delta = 2.5$, respectively.
In the bottom panel, we plot  $n(\bm{k})$
for the same two systems along a path in the Brillouin zone,  including the $\Gamma$ point $\bm{k} = (0,0)$, the antinodes $(0,\pi)$ and $(\pi,0)$, the node $(\pi/2,\pi/2)$ and the corner of the Brillouin zone $(\pi,\pi)$. We also show the momentum distribution of the corresponding half-filled systems, in order to probe the 
location of the excess holes in  $\bm{k}$-space.
At $\Delta = 4.4$, 
the momentum distribution appears to be smoother than at $\Delta = 2.5$, 
where the Fermi surface is much more defined and closer to the non-interacting structure.
This is consistent with the fact that
the system is more correlated at $\Delta = 4.4$, 
where more holes are on the $d$ orbitals, with  a higher number of double occupancies.
We observe a kink in the momentum distribution close to the antinodes, more
prominent at $\Delta = 4.4$, which reconstructs the Fermi surface 
from the non-interacting open diamond shape towards a closed circle.

\COMMENTED{
Interestingly, the comparison with the half-filled results suggests that 
the excess holes tend to occupy ``internal'' momenta close to $(\pi,\pi)$ and the node (and symmetry-related points)  in the system with $\Delta = 4.4$, while, for 
$\Delta=2.5$, they mostly occupy the antinodal and the nodal regions, close to the non-interacting Fermi surface. 
 }
 
 In Fig.~\ref{fig:mom_dist_pathbr}, we 
 show the corresponding \emph{band-resolved momentum distributions}. 
 We observe that the asymmetry between $p_x$ and $p_y$ orbitals can be understood as a consequence of the geometry of the lattice and the definition of the hopping amplitudes in the Hamiltonian in Eq.~\eqref{3bands:ham}. For a hole in the $p_x$ orbital, for example, it is more likely to have momentum in the $x$ direction, which is evident in Fig.~\ref{fig:mom_dist_pathbr}. 
 The comparison with the half-filled results in both Fig.~\ref{fig:mom_dist_both} and here provides a detailed picture of the behavior of the excess holes
 in momentum space.
 Upon doping, at $\Delta = 4.4$ the holes tend to occupy the $p$ orbitals close to $(\pi,\pi)$, while a percentage of them appear to occupy both $d$ and $p$ orbitals close to the the node $(\pi/2,\pi/2)$.
On the other hand, 
at $\Delta = 2.5$,  the excess holes appear to occupy $d$ and $p$ orbitals with momenta close to the antinode $(0,\pi)$, as well as close to the node $(\pi/2,\pi/2)$. 
The nesting that results from such arrangements clearly has to do with delicate spin orders we have observed.

\begin{figure}[ptb]
\includegraphics[width=\columnwidth, angle=0]{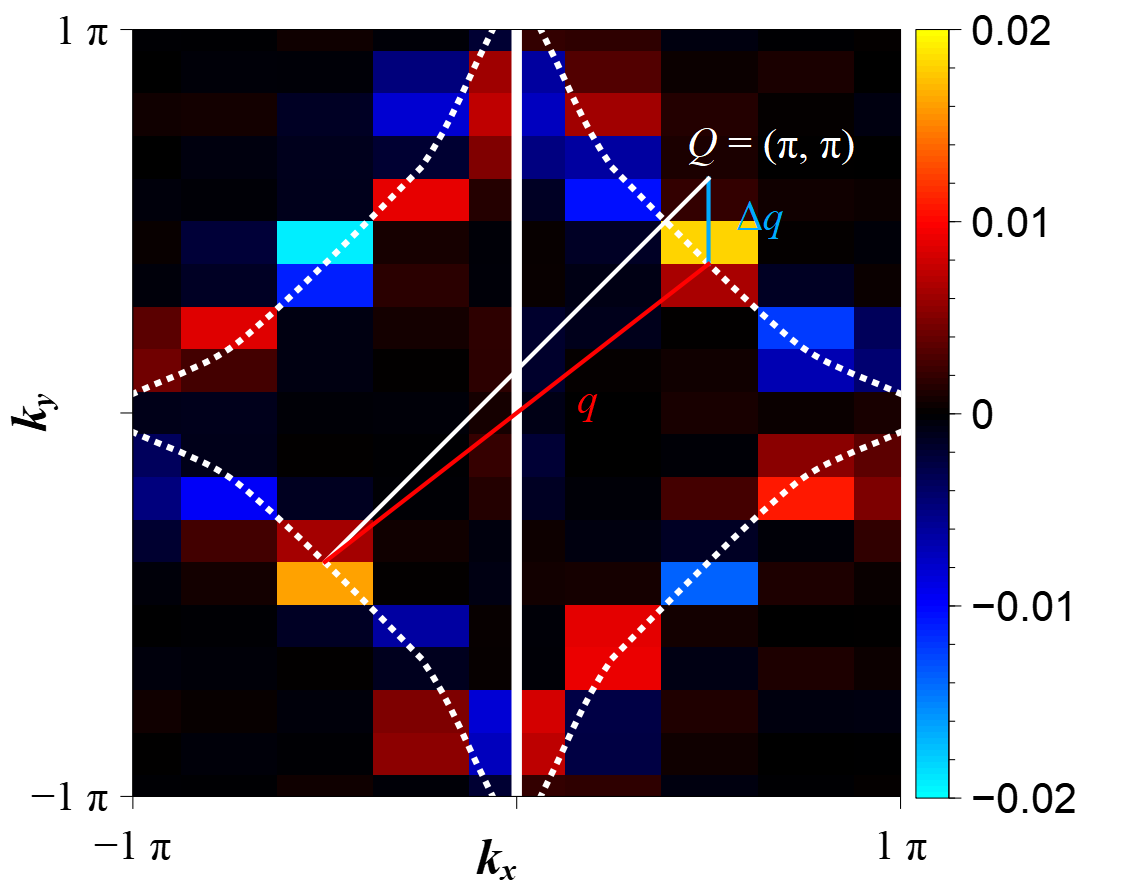}
\caption{
 (Color online) The difference between the spin-up and spin-down momentum distributions in the spiral state at $\Delta = 2.5$ with $h=1/8$.
 To guide the eye to complimentary nesting points on the Fermi surface,
 for $k_x < 0$, we plot $n_\uparrow (\bm{k})-n_\downarrow(\bm{k})$,
 and for $k_x > 0$, we plot $n_\downarrow (\bm{k})-n_\uparrow(\bm{k})$.
 We plot the nesting vector, $q$, in red, $Q = (\pi, \pi)$ in white, and $\Delta q = q-Q$ in blue.
 }
\label{fig:mom_diff_25}
\end{figure}

A remarkable difference between the two $\Delta$ values 
is seen in spin symmetry-breaking.
In the stripe-phase at $\Delta = 4.4$, the difference between $n_\uparrow (\bm{k})$ and $n_\downarrow (\bm{k})$ 
is negligible,  $n_\uparrow (\bm{k})-n_\downarrow (\bm{k})\sim 0$  within statistical error. 
In the spiral phase at $\Delta = 2.5$, $n_\uparrow (\bm{k})$ and $n_\downarrow (\bm{k})$ are not the same.
The difference $n_\uparrow (\bm{k}) - n_\downarrow (\bm{k})$  is crucial for the spiral order, 
as we extensively discussed at the mean-field level in [\onlinecite{Chiciak_GHF}].
In Fig.~\ref{fig:mom_diff_25}, we probe the differences between $n_\uparrow (\bm{k})$ and $n_\downarrow (\bm{k})$ 
at the many-body level.
Complimentary points are where $n_\uparrow (\bm{k})-n_\downarrow(\bm{k}) = n_\downarrow(\bm{k}^\prime)-n_\uparrow (\bm{k}^\prime)$. 
The vector connecting $\bm{k}$ and $\bm{k}^\prime$ is the nesting vector, $q$. 
We can then infer the difference ${\bf{\Delta q}} = {\bf{q}}-{\bf{Q}}$ between the spiral nesting vector 
${\bf{q}}$ and ${\bf{Q}}=(\pi,\pi)$ for the AFM order.  
The resulting ${\bf{\Delta q}}$ is along the $y$-direction, consistent with the
observed spiral state along $y$.
The resolution from QMC is  limited by the finite size of the system, 
in particular in the $x$ direction, such that it is difficult to infer ${\bf{q}}$ very precisely, 
but we estimate $\Delta q \simeq \pi/9$, which corresponds to a wavelength of 9 Cu sites in real space.
This is roughly consistent with the wavelength of 10 Cu sites discussed in the previous section.

\subsection{Localization of Holes}
\label{ssec:exotic}

In our QMC study of the Emery model at half--filling\cite{Ettore_three-band}, 
a phase transition was clearly identified between an AFM insulating state, 
which is stable at high values of $\Delta$, and a non--magnetic metal state which exists 
below $\Delta = 3.0$. One of the probes that we used to detect whether the system was insulating or conducting
was the 
Resta-Sorella localization estimator\cite{Resta-Sorella}.
Here we  also study  the localization of the holes  in the doped systems. 
However, since we have systematically used open boundary conditions, we will use as a probe 
the Quantum Metric Tensor (QMT) \cite{Resta-Sorella}, defined by the $2\times2$ matrix:
\begin{equation}
Q_{ab} = \frac{1}{N} \big( \left\langle \hat{r}_{a}  \hat{r}_{b}  \right\rangle - \left\langle \hat{r}_{a}  \right\rangle \left\langle  \hat{r}_{b}  \right\rangle \big), \quad  a, b = x, y
\end{equation}
The position operator is defined as:
\begin{equation}
\hat{r}_{a}  = \sum_{i=1}^{M} \sum_{\alpha = d, p_x, p_y} \left(r_{i,\alpha}\right)_a \, \sum_{\sigma}
\hat{\alpha}^{\dagger}_{i,\sigma} \hat{\alpha}^{}_{i,\sigma}\,,
\end{equation}
where $\left(r_{i,\alpha}\right)_a$ is the cartesian $a$-component of the position vector 
of the orbital $\alpha$ in the unit cell $i$. 
The diagonal components of the QMT 
provide a measure of the localization of the holes in the system.
In particular, since our supercells are elongated 
in the $y$ direction, we focus on the size dependence of the $Q_{yy}$-component of the QMT under open boundary condition. 
If $Q_{yy} \to \infty$ as $L_y \to \infty$, then we have a conductive state;
if $Q_{yy}$ converges to a finite value in the bulk limit, the system is an insulator.

\begin{figure}[ptb]
\includegraphics[width=\columnwidth, angle=0]{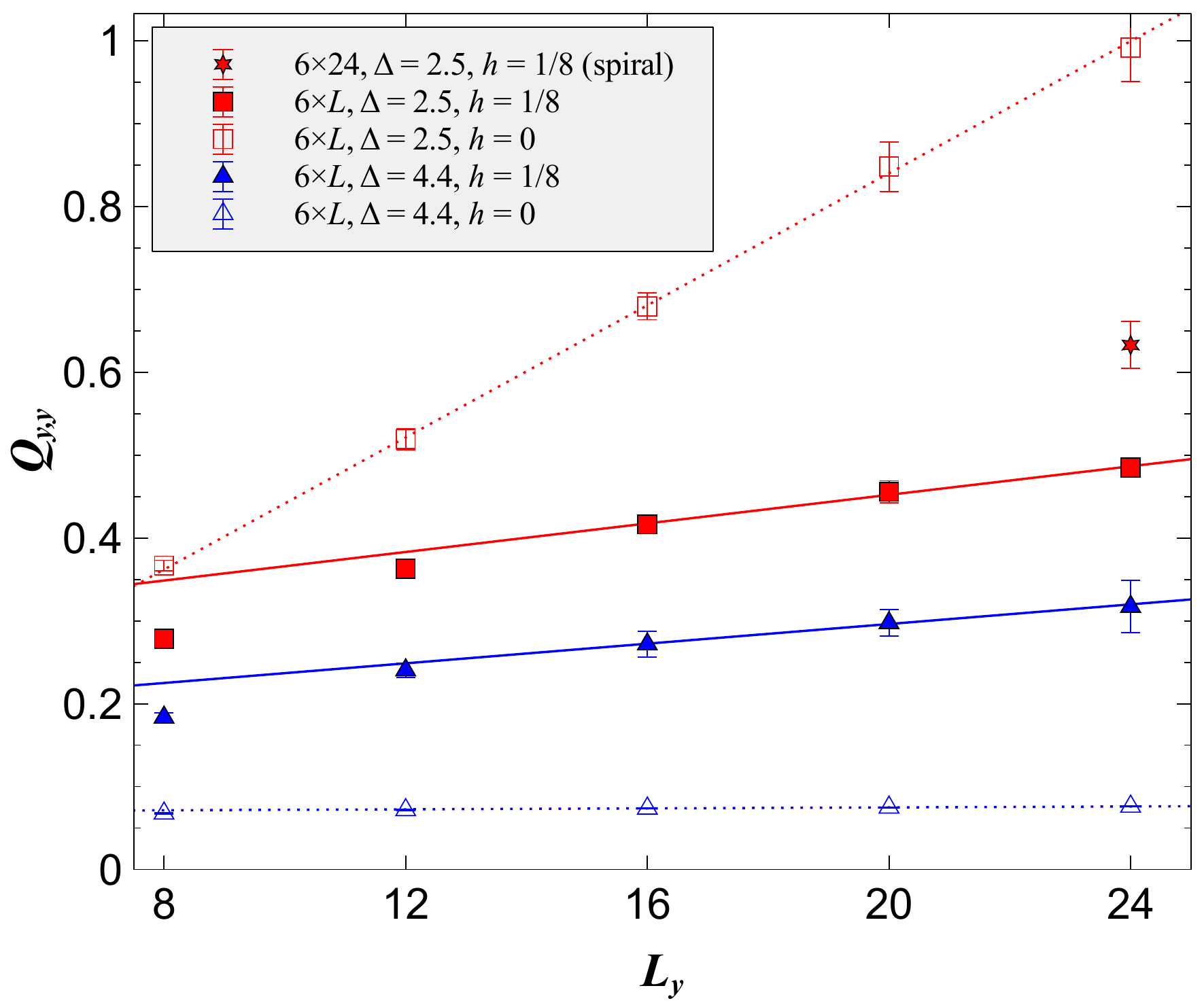}
\caption{
 (Color online) Plot of the $(y,y)$--component of the QMT as a function of
 $L_y$. At $\Delta = 4.4$ and half-filling, the value of the QMT is saturated for large 
 lattice sizes suggesting an insulating state. For both doped systems $\Delta = 2.5, 4.4$, the QMT appears to be 
 still increasing suggesting conducting states.
 }
\label{fig:localization}
\end{figure}

In Fig.~\ref{fig:localization}, we plot $Q_{yy}$ computed from AFQMC as 
a function of lattice size. 
We did not observe any significant difference between results for $L_x = 6$ and $L_x = 8$, 
indicating that the role of the transverse direction is negligible. For clarity and to maximize the length of the extrapolation, 
we only show results for 
$6 \times L_y$ systems.
The results at half-filling for both values of the charge-transfer energies are also shown for comparison, 
and provide a reference relating 
to our previous study \cite{Ettore_three-band}, which established 
that the ground state of the Emery model at half-filling 
is conductive at $\Delta = 2.5$  
and insulating at $\Delta = 4.4$. 

In the $1/8$-doped systems, the QMT 
increases as a function of the supercell size  for both values of $\Delta$. 
Interestingly, the slopes of $Q_{yy}$ as functions of $L_y$ for the two doped systems lie between the corresponding results at half-filling.
At $\Delta=2.5$, the excess holes appear to substantially reduce the overall mobility, but
the system still shows evidence of delocalized holes.
The mobility is substantially  higher in the spiral phase, which is not very surprising since the spiral order creates fewer (ideally no) domain walls, 
and less charge fluctuation, both of which should favor enhanced mobility. 
Interestingly, at $\Delta=4.4$ we see the opposite trend, with increased mobility upon doping.
The system seems to show signs of delocalized holes in the presence of stripe order, which is somewhat counter-intuitive. 
We stress that, although these system sizes are much larger than previously possible by accurate many-body computations,  we are still somewhat limited at   $L_y = 24$, especially for extrapolation of the asymptotic behavior. 
This makes it difficult to reach a conclusive answer about whether the ground state of the model is insulating or conductive.

As mentioned, we 
also computed the hopping amplitudes, namely the nearest-neighbor components 
of the one-body density matrix, as listed in Table~\ref{table:properties}.
These can be relevant to experiments, 
for example in scanning tunneling microscopy \cite{STM_Fischer}. 
The  matrix elements provide 
a further probe the local mobility of the holes. 
 From the results 
it is  evident that the local mobility of the holes increases 
as  $\Delta$ is decreased,
consistent with the QMT results above.

\subsection{Electron-Hole Asymmetry}
\label{ssec:asymmetry}

\begin{figure}[ptb]
\subfloat[$h=1/8$]{
\includegraphics[width=.4\columnwidth, angle=0]{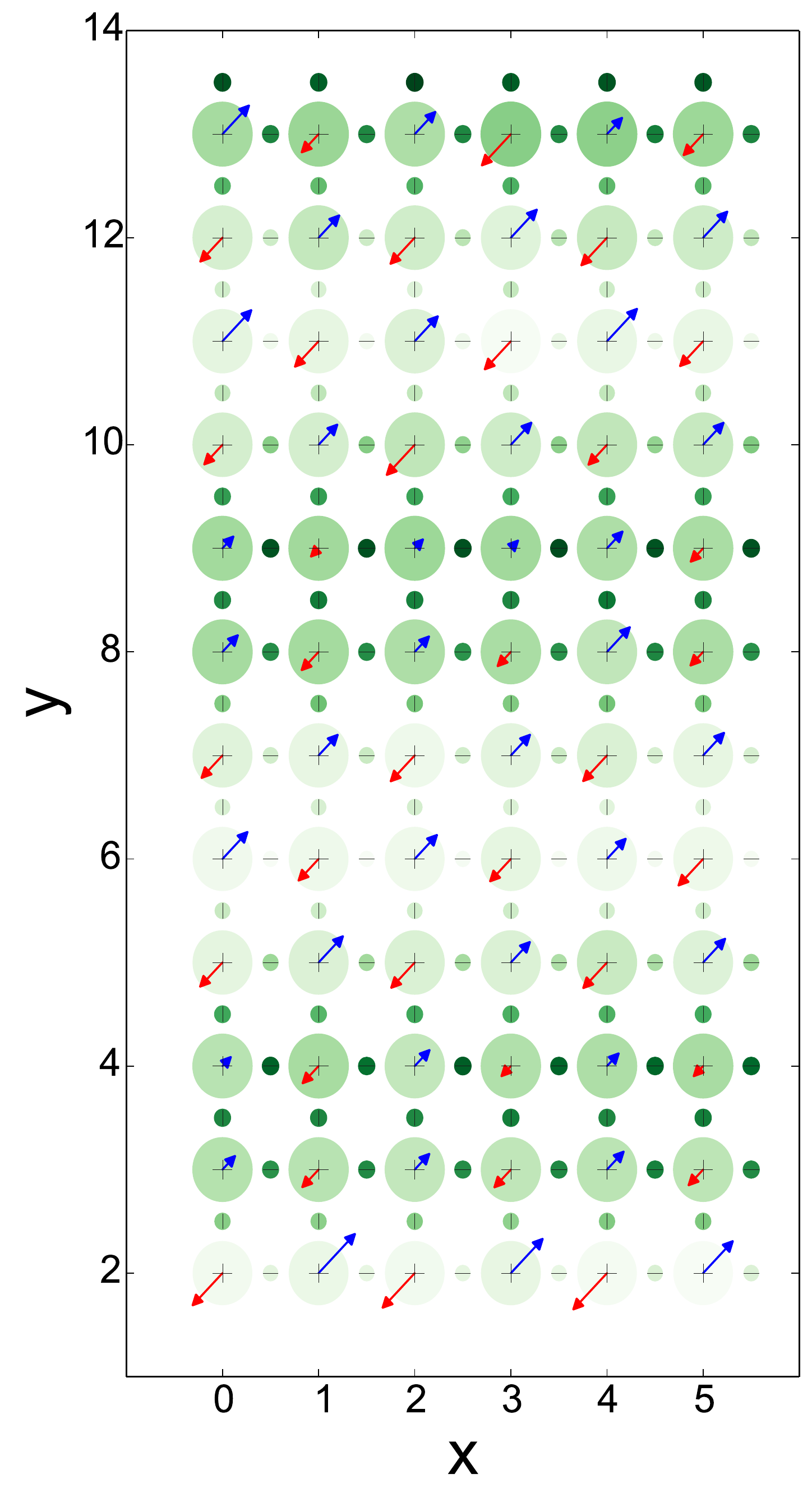}
}
\subfloat[$h=-1/8$]{
\includegraphics[width=.4\columnwidth, angle=0]{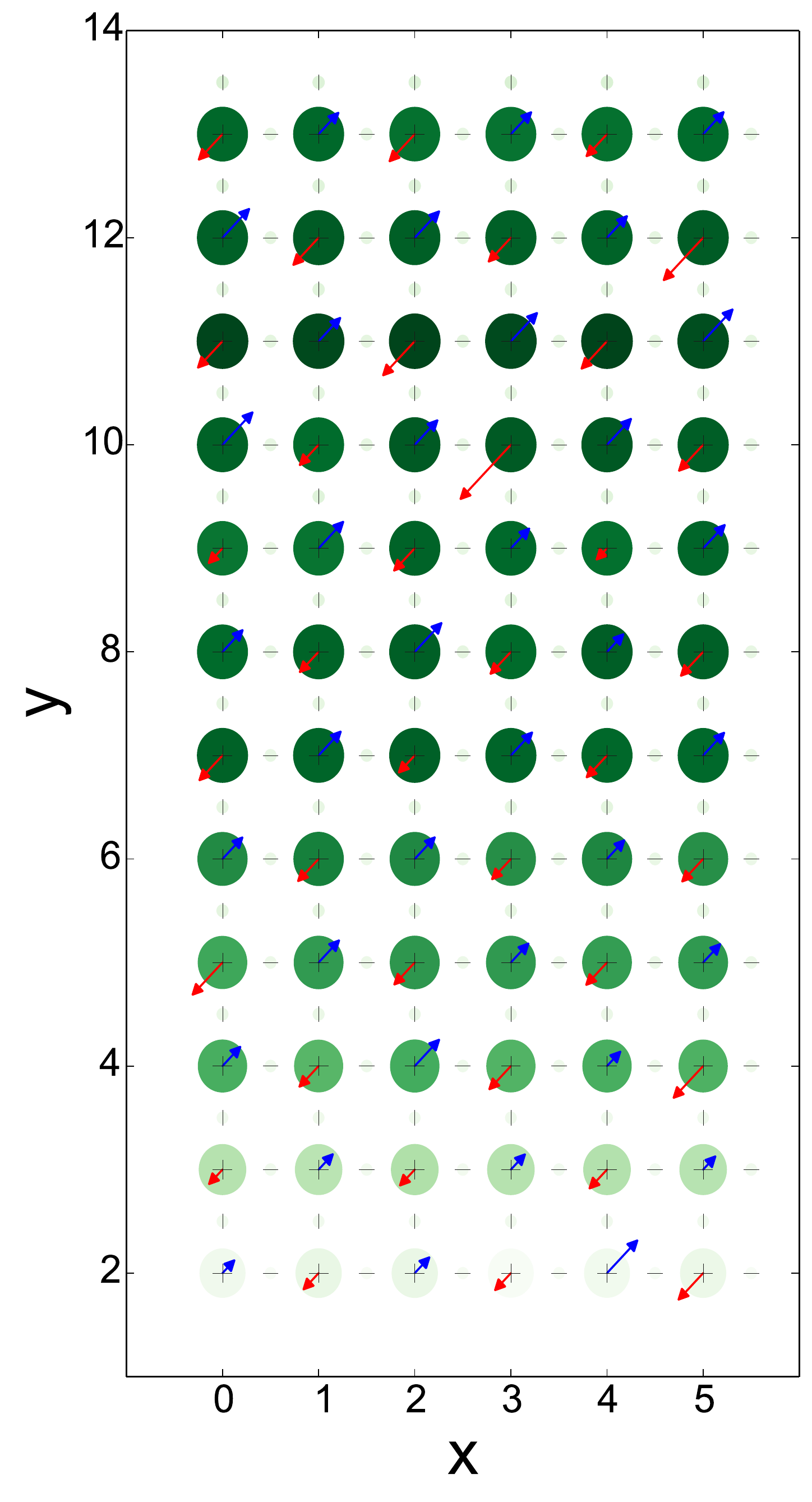}
}
\caption{
 (Color online) 
 Comparison of hole-doped and electron-doped systems.  
 The 2-D plots show  the $x$-component of the total spin, 
 $\langle\hat S_x({\mathbf r}) \rangle$, 
 and charge density,
 $ \langle\hat{n}({\mathbf r}) \rangle$, for both (a) a hole-doped system, $h=1/8$, and (b) an electron-doped system $h=-1/8$, at  $\Delta = 4.4$.
 The spins (arrows) are plotted as a projection in the $x$- plane. 
 The color of the arrow represents the 
 direction of  $\langle\hat S_x({\mathbf r}) \rangle$, 
 blue being positive and red negative. 
 The spin on the O $p$-orbitals is negligible and omitted from the plot.
 The size of the green circles are proportional to the hole density.
 The color of the circles is scaled to the maximum and minimum hole densities for the respective systems.
 The first and last two rows are not shown. 
 }
\label{fig:asymmetry}
\end{figure}

\begin{figure}[ptb]
\includegraphics[width=\columnwidth, angle=0]{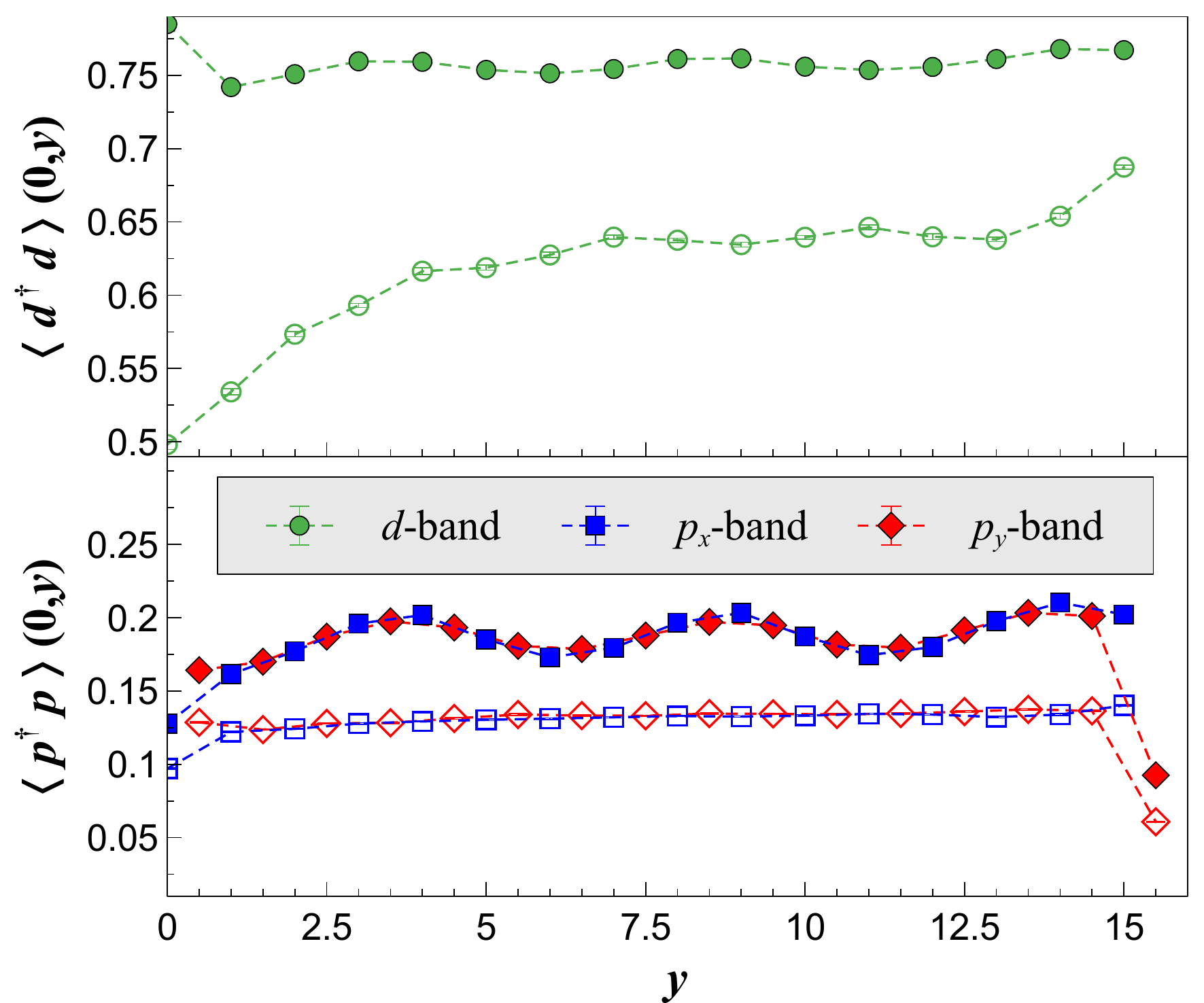}
\caption{
 (Color online) Plot of the occupations on different sites 
along the line
 cut at $x = 0$, for the systems in Fig.~\ref{fig:asymmetry}. The Cu $d$-band occupation is shown in the top panel, and 
 the O $p_x$ and $p_y$-bands are plotted in the bottom. Closed symbols represent the hole-doped system ($h=1/8$) and 
 open symbols represent the electron-doped system ($h=-1/8$).
 }
\label{fig:e-h_dens}
\end{figure}

An important feature of the cuprate phase diagram is the asymmetry between  electron-doping and hole-doping.
AFM correlations in the hole-doped case rapidly melt as holes are added to the system.
In the electron-doped case, on the other hand, the AFM 
state survives for much higher values of doping.
Although the main focus of this work is on the hole-doped regime, 
our methodology allows us to compute physical properties of electron-doped systems as well.
We have thus examined a system at $h=-1/8$ which mirrors one of the systems we have studied, in order to probe this asymmetry. 

In Fig.~\ref{fig:asymmetry}, we consider two $6\times16$ systems at $\Delta = 4.4$ in the three-band Hubbard model. On the left is a hole-doped system, 
exhibiting the behavior  consistent with what we have seen in Sec.~\ref{ssec:spin-charge}. 
The spin arrows help highlight the underlying AFM order, while the color variations of the charge circles highlight the density waves. 
a stripe phase in which the spin is modulated in phase with a charge wave.
On the right is an electron-doped system, with otherwise identical parameters. 
A strikingly different behavior is seen.
We observe a phase separated system in which a significant percentage of the doped electrons are localized on the $d$-orbitals
near the pinning line at $y = 0$. 
Beyond the inhomogeneous region induced by the pinning field, the systems aligns in a homogenous
AFM order. 

In Fig.~\ref{fig:e-h_dens} we show a quantitative comparison for the band-resolved hole density
between the two systems. 
It is evident that the majority of the doped electrons occupy $d$-orbitals, while doped holes tend to go to the $p$-orbitals with higher probability.
 In the electron-doped case, the system appears to phase separate in order to build an optimal density to form a homogenous AFM order, consistent with the experimental observation.
Our explorations in the electron-doped case are not as extensive as in the hole-doped case, where systematic computations for 
 different system sizes established the spin and charge order. It is possible that the AFM domains could acquire some modulation for larger system sizes.
 However, we tested  in supercells as large as $6 \times 24$ and it is clear 
  that such modulations would have much larger domain size than the wavelength in the hole-doped case, and they did not yield any obvious lowering of the energy compared to a state of a single domain.
The sharp  contrast between the electron- and hole-doped cases in the Emery model is an important step towards a more 
realistic model for the cuprates.



\section{Conclusions}
\label{sec:conclusion}

Using CP-AFQMC with the latest developments, we have studied  
the hole--doped, 
three--band Hubbard model as a function of the charge--transfer energy. 
 The magnetic and charge orders  are determined at two 
representative values of $\Delta$. 
Accurate numerical results are obtained from computations on large supercells to provide systematic information on 
a variety of ground-state properties.
Based on the performance of CP-AFQMC both in simplified models and in real materials,
these results represent the state-of-the-art in many-body computation for the combination of accuracy and approaching the 
bulk limit in the model.
Thus the detailed data will serve as useful benchmarks for future computational studies, as well as provide 
valuable cross-check for theoretical and experimental studies. 

Comparing the computed average Cu and O occupations to experimental studies, 
we find that,  with the parameters adopted,
the Emery model at $\Delta = 4.4$ most closely relates to the  Y-family of Cuprates, 
while at $\Delta = 2.5$, it 
most resembles the Hg-, Bi-, TI-based families.
At $\Delta = 4.4$, we observe a robust stripe order consisting of  spin density waves with 
corresponding 
charge density modulation, 
creating AFM background with 
a phase change across 
boundaries 
where the hole density in the vicinity is higher.
At $\Delta = 2.5$, on the other hand,  the spin order was more nuanced with several competing orders 
sensitive to the system sizes and geometries and initial trial wave functions. 
We find a spin density wave state, characterized by modulated AFM order along with a weak charge density wave only on the O $p$-sites, 
as well as a spin spiral state in which the spins cant in a randomly chosen plane along the propagation direction with essentially uniform charge density.
These states are separated by an energy scale that is almost degenerate within the 
(high) resolution of the AFQMC calculation,
suggesting a possible quasi-degeneracy of the ground state of the Emery model. 

We characterized the properties of these states with detailed information on the densities in supercells with a pinning field applied on one side to 
break translational symmetry. We also computed average hopping amplitudes and energetics as detailed in  Table~\ref{table:properties}. 
The momentum  distributions were analyzed and compared for the stripe and spiral states. 
We observed that the holes became more delocalized
as the charge--transfer energy was reduced, by measurements of the QMT 
and the one--body density matrix. The spiral spin state, which has a nearly constant charge density, has holes substantially more delocalized than in the stripe state.  
Finally, we explored the relation between hole- and electron-doping and found that the Emery model exhibited an asymmetry in the AFM orders off half-filling which is consistent with the observed phase diagrams of cuprate materials.

The Emery model shows significant differences from the one-band Hubbard model at the mean field level.
The ground state from generalized Hartree-Fock exhibits  \cite{Chiciak_GHF} a very  rich phase diagram 
including orders such as diagonal magnetic domain walls, nematicity, and spin spirals.
At the many-body level, some of these features from GHF were not observed. 
At $\Delta = 4.4$  the half-filled system has AFM order and is insulating, while
the $1/8$-hole-doped system exhibits a stripe order 
rather similar to what is 
seen in the one-band model. The spiral state at  $\Delta = 2.5$, which is either the ground state or nearly degenerate 
with an SDW ground state, has not been seen in the simple one-band Hubbard model. (It is not clear whether some engineering 
of the hopping parameters beyond near-neighbors will make this state also appear in the one-band model.) 
Based on these results one is tempted to reinforce that the three-band model is perhaps only marginally more relevant than the
one-band Hubbard model for representing the cuprates.
However, 
the answer is more nuanced regarding how similar the Emery model is to the one-band Hubbard model.

 The Emery model captures the asymmetry in AFM order between hole-doping and electron-doping  seen in the phase diagram of the real materials, which is not present in the particle-hole symmetric one-band model.
As we showed, the model with different values of  $\Delta$  reproduced, to an excellent degree for  different families of cuprates, the
experimentally measured $d$ and $p$ orbital occupancies, $n_d$ and $n_p$, which are known to affect several properties including 
 the superconducting transition temperature.
Results for  the fate of excess holes and the localization also appear to mirror well the phenomenology of the different families of real materials. 
The ground-state properties show considerable 
sensitivity to parameter values and details. This basic feature is seen even in the one-band model, and is more pronounced in the Emery model, as 
reflected both in the variation with $\Delta$ and in the delicate balance at $\Delta = 2.5$ that we have observed. 
Indeed the presence of many competing or cooperating orders within small energy windows is a trademark of the real materials 
whose essential physics we hope to capture with these models. 
It is thus reasonable to assume, especially without precise knowledge of what balance of these 
states would be responsible for superconductivity, that the Emery model can be different in a non-trivial way.

A major remaining question about the ground-state of the Emery model is of course superconductivity. 
We have not studied the nature of superconducting correlations in this work. Since our computations were done in the canonical ensemble,
we could not directly measure superconducting order parameter. Pairing correlation functions can be measured, however these will have 
very small amplitude and will require systematic finite-size scaling with high resolution to determine the asymptotic (distance) behavior unambiguously.
Recent progress in the one-band model  \cite{Mingpu-pure-Hubbard-SC} suggests a variation in AFQMC which provides a
promising avenue to determine pairing order.
We will leave this to a future investigation.

We thank the Simons Foundation for support. Computing was performed using resources from 
XSEDE,
which is supported by National Science Foundation grant number ACI-1053575, 
 and the OLCF at ORNL which is supported by the Office of Science of the U.S. Department of Energy under contract no.~DE-AC05- 00OR22725.
We also acknowledge the High Performance Computing at William \& Mary for their resources and help.
We thank Andrew Millis, 
Henry Krakauer, Enrico Rossi, Hao Shi, Mingpu Qin, and Hao Xu for useful feedback and conversations,
and Lucas Wagner for providing us with the parameter values from Table~\ref{tab:param_table}.
The Flatiron Institute is a division of the Simons Foundation.


\input{paper.bbl}

\end{document}

%% file: paper.bbl
%